\documentclass[aps,prb,twocolumn, superscriptaddress]{revtex4-1}
\usepackage{amsmath}
\usepackage{epsfig}
\usepackage{graphicx}
\usepackage{multirow}
\usepackage{array}
\usepackage{inputenc}
\usepackage{bm}
\usepackage{color}
\usepackage{gensymb}

\newcommand{\hoti}{Ho$_2$Ti$_2$O$_7$}
\newcommand{\dyti}{Dy$_2$Ti$_2$O$_7$}
\newcommand{\tbti}{Tb$_2$Ti$_2$O$_7$}
\newcommand{\ybti}{Yb$_2$Ti$_2$O$_7$}

\newcommand{\yb}{Yb$^{3+}$}

\begin{document}
\author{J. Robert}\email[]{julien.robert@neel.cnrs.fr}
\affiliation{Laboratoire L\'eon Brillouin, CEA-CNRS UMR 12, Centre de Saclay, F-91191 Gif-sur-Yvette, France}
\affiliation{Institut N\'eel, CNRS \& Universit\'e Joseph Fourier, BP 166, 38042 Grenoble Cedex 9, France.}
\author{E. Lhotel}\affiliation{Institut N\'eel, CNRS \& Universit\'e Joseph Fourier, BP 166, 38042 Grenoble Cedex 9, France.}
\author{G. Remenyi}\affiliation{Institut N\'eel, CNRS \& Universit\'e Joseph Fourier, BP 166, 38042 Grenoble Cedex 9, France.}
\author{S. Sahling}\affiliation{Institut N\'eel, CNRS \& Universit\'e Joseph Fourier, BP 166, 38042 Grenoble Cedex 9, France.}
\affiliation{TU Dresden, Institut f\"ur Festk\"orperphysik, D-01062, Germany}
\author{I. Mirebeau}\affiliation{Laboratoire L\'eon Brillouin, CEA-CNRS UMR 12, Centre de Saclay, F-91191 Gif-sur-Yvette, France}
\author{C. Decorse} \affiliation{LPCES, Universit\'e Paris-Sud, 91405, Orsay, France}
\author{B. Canals}\affiliation{Institut N\'eel, CNRS \& Universit\'e Joseph Fourier, BP 166, 38042 Grenoble Cedex 9, France.}
\author{S. Petit}\affiliation{Laboratoire L\'eon Brillouin, CEA-CNRS UMR 12, Centre de Saclay, F-91191 Gif-sur-Yvette, France}

\title{Spin dynamics in the presence of competing ferro- and antiferro-magnetic correlations in \ybti}
\date{\today}
\begin{abstract}
In this work, we show that the zero field excitation spectra in the quantum spin ice candidate pyrochlore compound \ybti\ is a continuum characterized by a very broad and almost flat dynamical response which extends up to $1-1.5$~meV, coexisting or not with a quasi-elastic response depending on the wave-vector. The spectra do not evolve between 50~mK and 2~K, indicating that the spin dynamics is only little affected by the temperature in both the short-range correlated and ordered regimes. Although classical spin dynamics simulations qualitatively capture some of the experimental observations, we show that they fail to reproduce this broad continuum. In particular, the simulations predict an energy scale twice smaller than the experimental observations. This analysis is based on a careful determination of the exchange couplings, able to reproduce both the zero field diffuse scattering and the spin wave spectrum rising in the field polarized state. According to this analysis, \ybti\ lies at the border between a ferro and an antiferromagnetic phase. These results suggest that the unconventional ground state of \ybti\ is governed by strong quantum fluctuations arising from the competition between those phases. The observed spectra may correspond to a continuum of deconfined spinons as expected in quantum spin liquids. 
\end{abstract}

\maketitle


\section{Introduction}
Understanding, characterizing and classifying novel states of matter is one of the main goal of the research in solid state physics. In particular, systems where the thermal or quantum fluctuations are able to melt long range order, the so-called spin liquids, draw a lot of attention since they generally go beyond N\'eel paradigm \cite{Lacroix}.

The pyrochlore lattice, made of corner-sharing tetrahedra is the archetype of a three-dimensional frustrated lattice and has proven during the last years to be a rich playground for studying such spin liquid states \cite{gingrasrmp}. 
Among the variety of possible pyrochlore systems, the compound \ybti\, has been presented as one of the possible realizations of a quantum spin liquid ground-state. More precisely, it has been proposed as a candidate for the quantum variant of the spin ice state observed in \hoti\, and \dyti. In these classical spin ices, the Ising-like anisotropy of the magnetic moments along the local $\langle111\rangle$ directions, combined with an effective ferromagnetic interaction induce a macroscopically degenerated ground state characterized by the local ice-rule \cite{harris} (in each tetrahedron, two spins point in and two spins point out). 
In \ybti, the \yb\ magnetic moment shows a weak XXZ planar anisotropy perpendicular to the local $\langle111\rangle$ directions \cite{hodges01, malkin04, cao1}. However, the strongly anisotropic interactions tensor \cite{cao2,thompson,rossprx}, whose main component is ferromagnetic, induces a local constraint analog to the ice-rule.

The static and dynamical magnetic correlations in \ybti\, have been investigated by means of neutron scattering experiments. Short range 
correlations settle around $T_0\sim 2$~K, giving rise to rods of diffuse scattering along high symmetry directions \cite{ross1, thompson, ross2}. At lower temperature, a phase transition is observed in specific heat measurements around 0.2 K \cite{blote}, but the critical temperature depends on the nature of the samples (polycrystal or single crystal) and their quality \cite{ross2,yaouanc,dortenzio}. This transition was shown to be first order \cite{hodges02, elsa, chang, dortenzio}, in both single crystals and powder samples but its nature remains debated. Several studies (including neutron scattering and muon spin relaxation measurements) evidence the stabilization of an ordered ferromagnetic state, with a reduced static magnetic moment \cite{chang, yasui, hodges02, elsa}, while others do not \cite{ross1, dortenzio, gardner04, bonville}.

To understand the underlying microscopic mechanism leading to such a non-conventional ground-state, several studies attempted to determine the exchange couplings, trying to reproduce the experimental results. Different sets of interactions were obtained, depending on the fitted quantity: diffuse scattering \cite{thompson, chang} or field induced spin wave excitations \cite{rossprx}. The latter parameters allow reproducing specific heat \cite{applegate}, and were further refined using terahertz spectroscopy measurements \cite{pan}. These parameters place \ybti\ in a "splayed ferromagnetic" phase, quite far from the "quantum spin liquid" phase of the theoretical phase diagram \cite{savary,wong,ludo}.

Since the precise nature of the elementary excitations generally depends on the ground-state itself, we propose in this article to address the unconventional magnetic properties of \ybti\ through a detailed study of the spin dynamics on a single crystal in the zero field correlated phase below 2~K, which has been little explored up to now \cite{ross2}. 
%
In section \ref{sec:exp}, we first present macroscopic measurements performed on a piece of our single crystal, which show evidence for a transition at 175 mK. By means of inelastic neutron scattering, we show that the excitations spectrum is a continuum characterized by a very broad and almost flat dynamical response extending up to $1-1.5$~meV, and coexisting or not with a quasi-elastic response depending on the wave-vector. These excitations do not depend on the temperature below the stabilization of short-range correlations (i.e. 2 K), and, in particular, entering the ordered phase does not affect the spectra. Then, in section \ref{sec:num}, we compare these inelastic data with classical calculations combining Monte Carlo and spin dynamics simulations. We show that, although the qualitative features of the spectra are reproduced, calculations predict an energy scale twice smaller than the experimental observations. 
These calculations are based on our determination of a reliable and robust set of exchange couplings, able to reproduce both the spin-wave spectrum of the field polarized phase as well as the diffuse scattering in zero field.
The obtained couplings do place \ybti\ in a ferromagnetic phase, yet very close to the boundary with an antiferromagnetic phase. 
Then, although \ybti\ is not a canonical quantum spin liquid as initially proposed, we propose that the proximity of competing ferromagnetic and antiferromagnetic correlations restore strong quantum fluctuations\cite{ludo} and is at the origin of the anomalous static and dynamical behaviors observed in our experiments.

\section{Experimental}
\label{sec:exp}

Our measurements were performed on a single crystal. First, a polycristalline sample of \ybti\ stoichiometry was synthesized from Yb$_2$O$_3$ and TiO$_2$ starting powders by solid state reaction. To obtain the pyrochlore phase several thermal treatments at 1400 \degree C with intermediate grindings were necessary. The progress of the synthesis reaction was followed by powder X-ray diffraction. For the last thermal treatment the powder was shaped as a cylinder of 5 mm diameter and 90 mm length and the obtained road was used as feed road for the single crystal synthesis. Crystal growth was performed using the floating zone technique in a four-mirror optical image furnace NEC SC-N15HD. The obtained crystal was then annealed under O$_2$ gas flow for two days. 

\subsection{Thermodynamic measurements: Magnetization and specific heat}
\label{sec:exp_mag}
To characterize the macroscopic properties of our \ybti\ sample, we have performed thermodynamic measurements, magnetization and specific heat, on a flat disk sample (mass 244 mg), cut from the single crystal used for neutron scattering experiments. These measurements were performed at the Institut N\'eel in purpose built experiments equipped with $^3$He-$^4$He dilution refrigerators. Magnetization was measured in a SQUID magnetometer as a function of field and temperature down to 90 mK, together with ac susceptibility measurements \cite{Paulsen01}. Specific heat was measured by the relaxation method down to 70 mK in several applied fields up to 100 mT. The high sensitivity and fast response of the thermometers used in this experimental set-up as well as their high stability allow to measure both short and long time heat relaxation \cite{lasjaunias}. The magnetic field was applied in an arbitrary direction, parallel to the plane of the disk. 

\begin{figure}
\includegraphics[keepaspectratio=1, width=8.5cm]{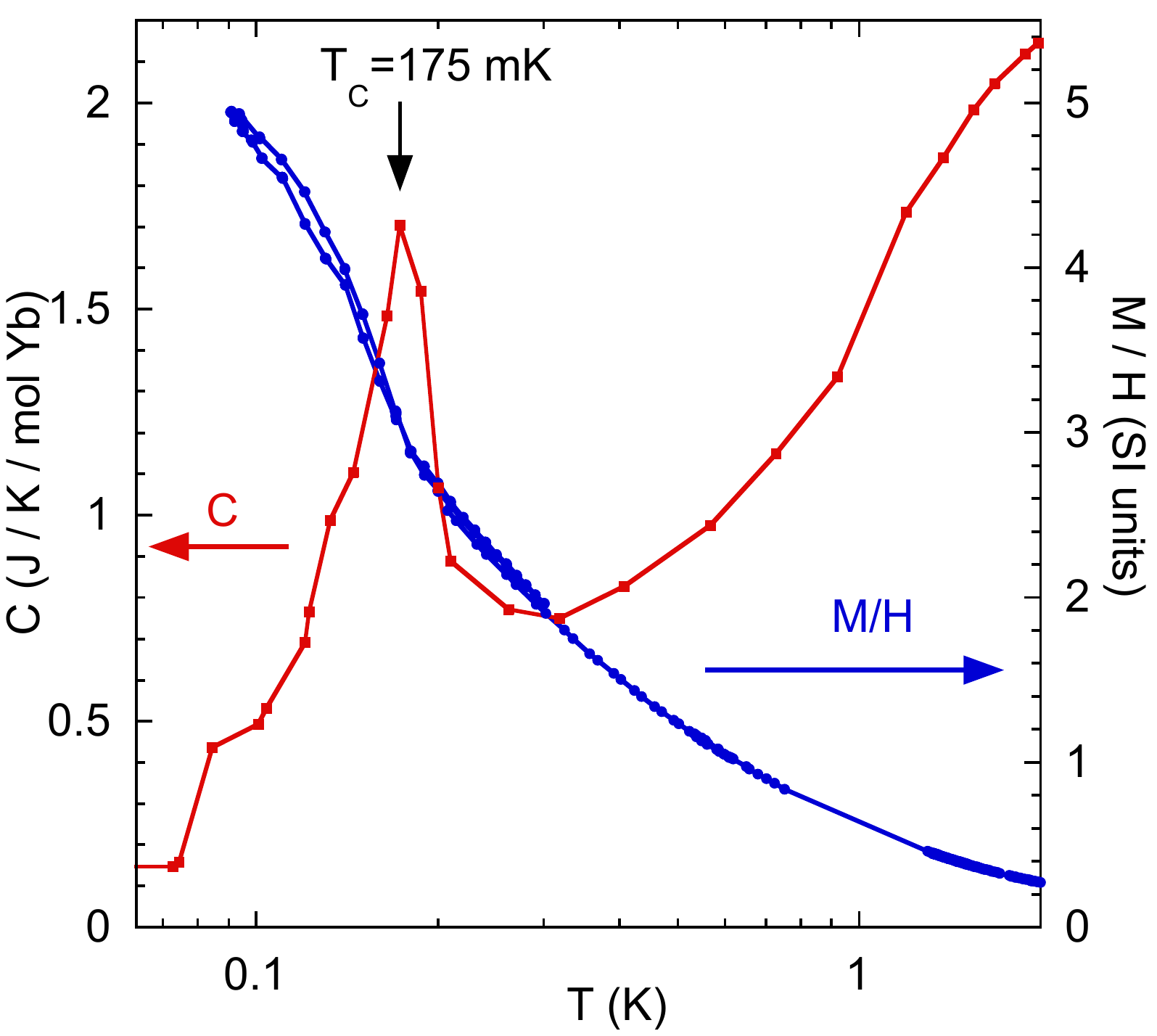}
\caption{(Color online) Specific heat $C$ (red squares) and magnetization $M/H$ (blue dots) vs temperature $T$. The specific heat is the long time specific heat (see text) and the magnetization was measured in a field of 3.83 mT, with a step of 10 mK every 1350 s.}
\label{fig_CMvsT}
\end{figure}

The magnetization exhibits a Curie-Weiss temperature of 0.5 K, and reaches 1.65 $\mu_B$/Yb at 90 mK and 3 T which is consistent with previous measurements \cite{elsa, hodges01, yasui, bramwell, thompsonjpcm}. The specific heat presents a broad maximum around 2.5 K characteristic of short-range order correlations in agreement with previous results \cite{blote}. When decreasing the temperature, these measurements show a transition at $T_C=175$ mK, in both specific heat and magnetization, which is evidenced by a peak in the specific heat, an upturn in the magnetization (see Figure \ref{fig_CMvsT}), and the onset of the out-of-phase ac susceptibility (not shown). As previously observed \cite{elsa}, the magnetization shows a small hysteresis between the cooling and warming curves below the transition, indicative of a first order transition. Nevertheless, although the peak in specific heat is rather narrow, the increase of the magnetization when decreasing the temperature below the transition is smooth, and fails to saturate, even at 90 mK. This result suggests that the magnetic transition expands on a broad temperature range and that no net spontaneous ferromagnetic moment is stabilized in this sample down to 90 mK. 

\begin{figure}
\includegraphics[keepaspectratio=1, width=7.5cm]{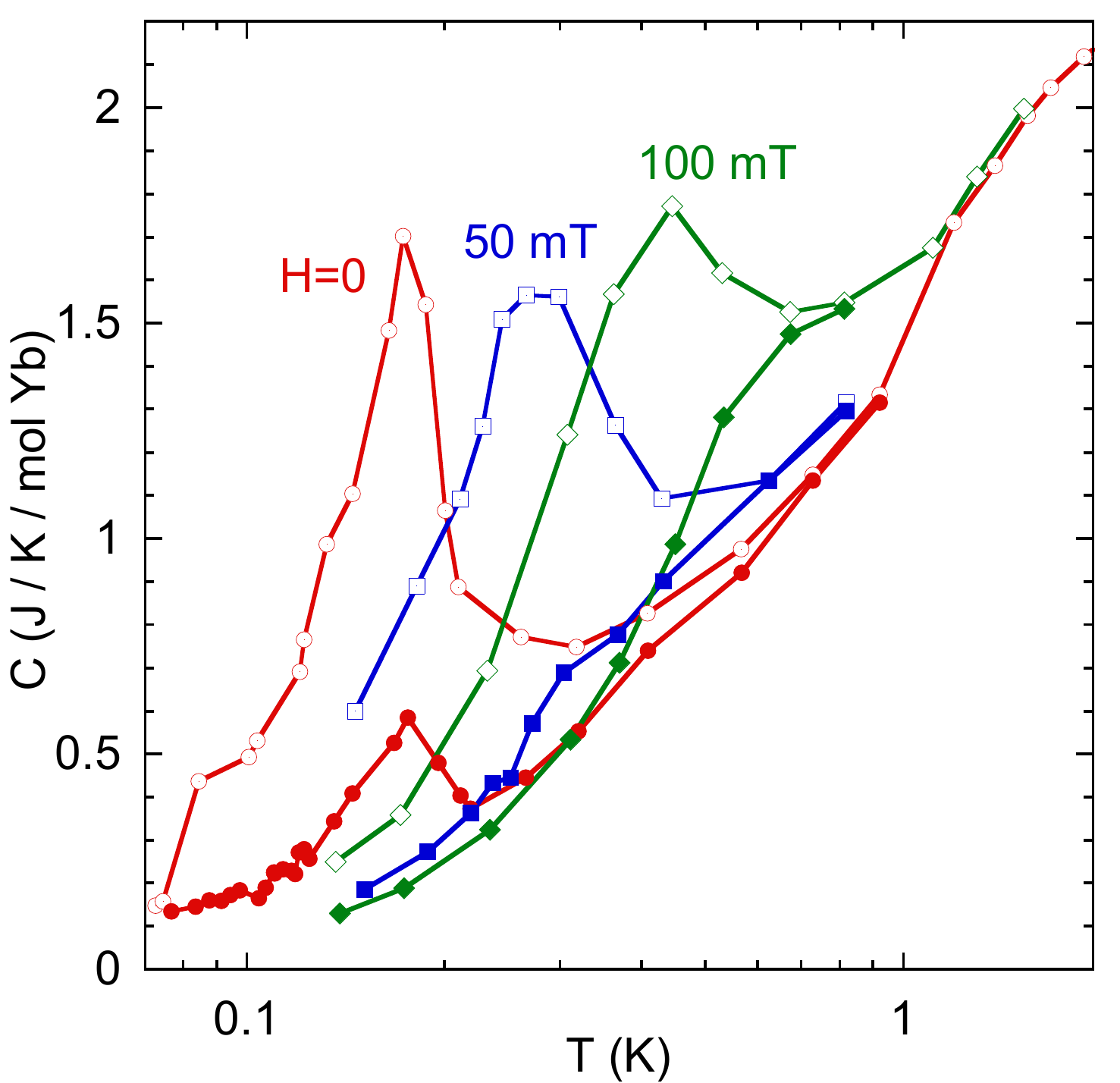}
\caption{(Color online) Specific heat $C$ vs temperature $T$ for several magnetic fields. Open symbols correspond to the long-time specific heat ($t \approx 600$ s) and full symbols to the short time specific heat ($t \approx 9$ s). }
\label{fig_CvsH}
\end{figure}

When a small field is applied ($\mu_0 H \leq 100$ mT), the transition persists and is shifted to higher temperatures, confirming the ferromagnetic nature of the transition (see Figure \ref{fig_CvsH} for specific heat). It is worth noting that below 1 K, the heat relaxation is not exponential, resulting in two contributions for the specific heat: the short time one (characteristic time of about 9 s) and the long time one (characteristic time of about 600 s). These contributions show qualitatively the same temperature dependence in zero magnetic field, which evidences that dynamics with different time scales exist in the system above and below the transition. When increasing the magnetic field, the peak in the short time response is suppressed more drastically than the long one (See Figure \ref{fig_CvsH}). 

\subsection{Neutron scattering}
\label{sec:exp_neutrons}

The neutron scattering measurements were performed on a large \ybti\, single crystal grown with the floating zone technique, as detailed above. The crystal was aligned in the $(hh0) -(00\ell)$ scattering plane and cooled down to 50~mK in a dilution fridge. Note that due to the large size of the crystal, its poor thermal conductivity at very low temperature, and the heat provided by the neutron beam, the true temperature of the crystal might be slightly larger. Special care to the thermalisation of the sample was taken during the experiments: a waiting time of a few hours was used between the measurements to ensure a stable temperature inside the sample. 
Polarized and unpolarized experiments were respectively conducted on the 4F1 and 4F2 triple-axis spectrometers (LLB, France), with final neutron wave-vectors $k_f = $1.3 $\AA^{-1}$, resp. $k_f = $1.15 $\AA^{-1}$, yielding an energy resolution of $\Delta_r = $ 150 $\mu$eV, resp. $\Delta_r = $ 85 $\mu$eV. A cooled beryllium filter was placed in the incident beam to remove high order contaminations. The polarization analysis allows measuring the spin-spin correlation functions $S_y(\mathbf{Q},\omega)$ and $S_z(\mathbf{Q},\omega)$ where the subscripts $y$ and $z$ indicate that the spin components are perpendicular to $\mathbf{Q}$ within the scattering plane and along the vertical axis respectively. Both are measured in the spin flip channel, with polarization (direction of the incident neutron's spin) $\mathbf{P}$ applied along $z$ and $y$ respectively. The flipping ratio was about FR=42 at high temperature but slightly decreased below $T=1$~K, down to FR=33 at $T=0.65$~K. Because of technical difficulties during the polarized neutron experiment, the lowest reachable temperature was $T\simeq 650$~mK. Note that the magnetic intensity measured in unpolarized experiments is the usual spin-spin correlation function $S(\mathbf{Q},\omega) = S_y(\mathbf{Q},\omega)+S_z(\mathbf{Q},\omega)$.

\subsubsection{Static properties}
\label{sec:exp_ela}

\begin{figure}
\includegraphics[keepaspectratio=1, width=8.5cm]{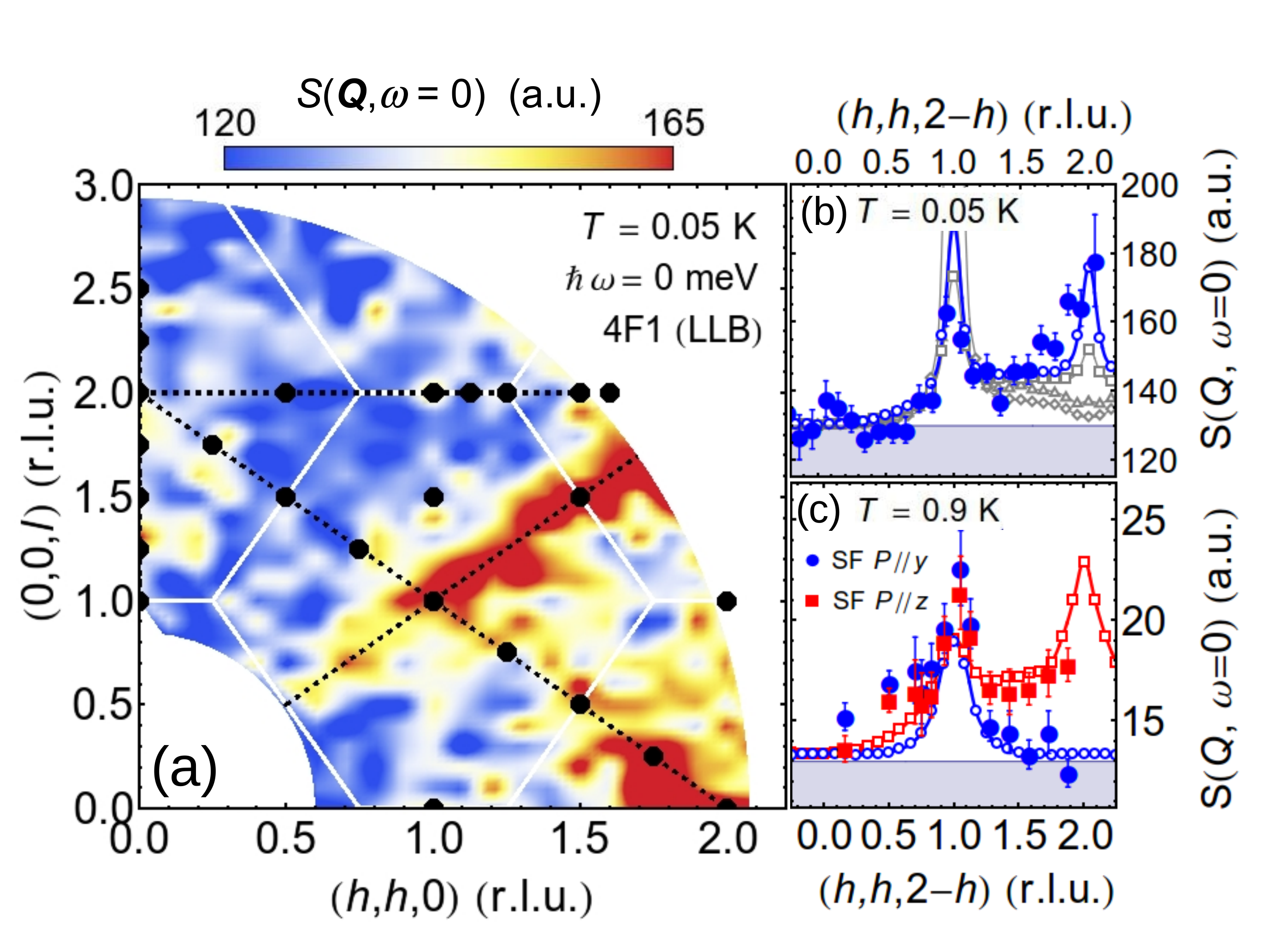}
\caption{(Color online) (a) Elastic intensity map in the $(h~h~\ell)$ scattering plane at $T=0.05$~K. White lines correspond to the Brillouin zones boundaries. Black dashed lines denote the directions probed in the inelastic neutron scattering experiments (see section \ref{sec:exp_ine}). (b) Elastic intensity (full dots) along the $(h~h~2-h)$ direction obtained by integrating Figure (a) over $\delta Q=0.4$~rlu in the perpendicular direction. The lines with open symbols show calculated elastic scattering functions obtained from classical spin dynamics simulations for different parameter sets $J_2=-0.26,-0.29,-0.31$~meV (gray), $-0.326$~meV (blue) and $T=0.4$~K (see section \ref{ssec:simus}). (c) Spin flip elastic intensities (full symbols) along the $(h~h~2-h)$ direction at $T=0.9$~K with polarization applied along $y$ (blue dots) and $z$ (red squares). The blue and red lines with open symbols correspond to calculated elastic scattering functions $S_y(\mathbf{Q},\omega=0)$ and $S_z(\mathbf{Q},\omega=0)$ obtained from classical spin dynamics simulations for $J_2=-0.32$~meV (see section \ref{ssec:simus}).}
\label{fig1:elastic}
\end{figure}

Elastic data collected at $T=0.05$~K are shown in Figure \ref{fig1:elastic}, revealing the same qualitative features as the energy integrated diffuse scattering described at length in prior works above the critical temperature \cite{ross2,thompson,chang, bonville}. It is characterized by a line of scattering along $(h~h~h)$ accompanied by another branch from $\mathbf{Q}=(1,1,1)$ to $(2,2,0)$ ending with a large spot at $\mathbf{Q}=(2,2,0)$. The presence of spectral weight around both $(1,1,1)$ and $(2,2,0)$ is better evidenced in Figure \ref{fig1:elastic} (b), where the elastic intensity is plotted versus wave-vector along the $(h~h~2-h)$ direction. 
While these rods of diffuse scattering are undoubtedly related to the presence of strongly anisotropic exchange interactions \cite{thompson}, their origin is still unclear. First ascribed to a dimensionality reduction of the static correlations \cite{ross2}, the ground-state has more recently been shown to be fully 3D in the region of parameter space where \ybti\, is supposed to lie \cite{ludo}.

The spectral weight along the rods is particularly intense around the $(2,2,2)$ Brillouin zone center, as expected in the presence of ferromagnetic correlations : in the limit of a long range collinear ferromagnetic order, the strongest intensity is obtained for $(2n,2n,2n)$ Bragg peaks, while the $(2n+1,2n+1,2n+1)$ ones are around three times weaker and the $(2n,2n,0)$ peaks are extinct.
A finite spectral weight at $\bm Q = (2,2,0)$ is characteristic of antiferromagnetic spin arrangements, which points out the coexistence of both ferro- and antiferromagnetic correlations in the system. 

These results reveal that the short range correlations still remain at very low temperature and coexist with the long range ferromagnetic order. This might be related to the fact that, in our sample, the magnetization increases smoothly below the transition temperature.

Finally Figure \ref{fig1:elastic} (c) displays data in the same direction measured using polarized neutrons at higher temperature ($T=0.9$~K). It points out that the signal around $\mathbf{Q}=(2,2,0)$ is polarized along $z$, while the one around $(1,1,1)$ has similar intensity in both the $y$ and $z$ channels. The present data are thus consistent with the one obtained in Ref. \onlinecite{chang}.

\subsubsection{Dynamic properties}
\label{sec:exp_ine}

\begin{figure}
\centerline{
\includegraphics[keepaspectratio=1, width=8.5cm]{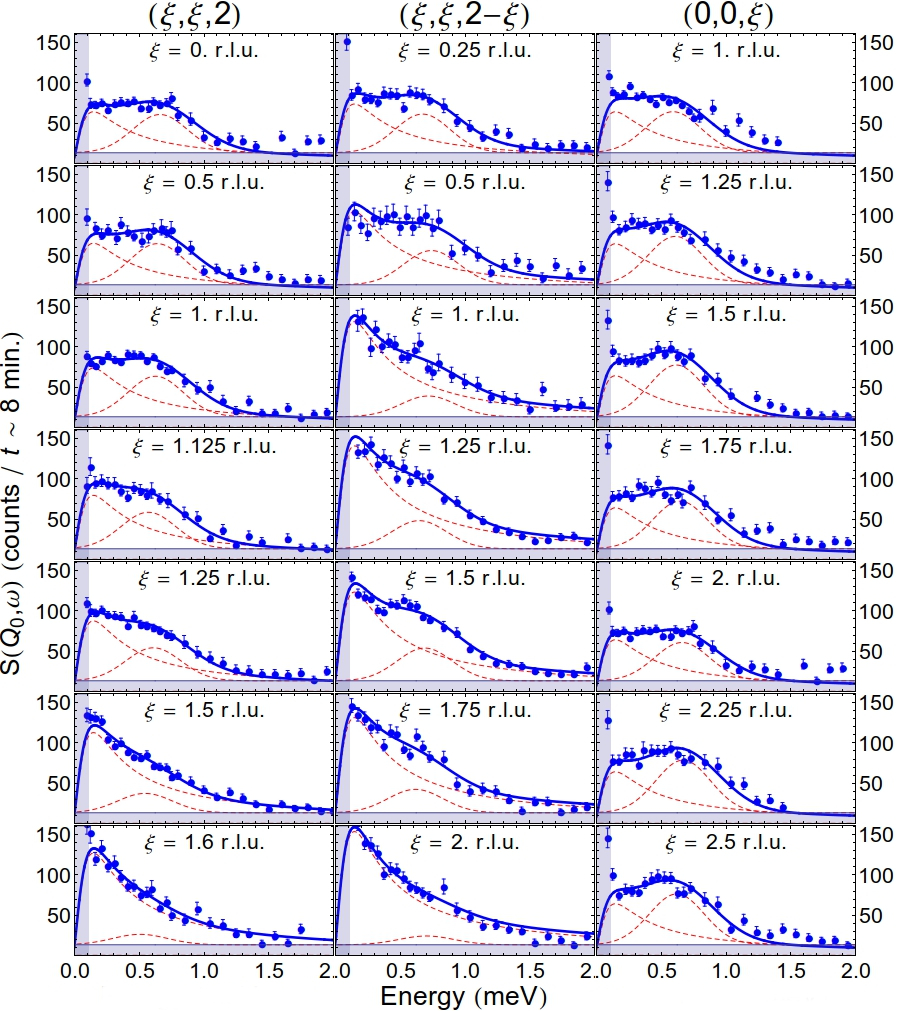}
}
\caption{(Color online) Excitation spectra along high symmetry directions $(h~h~2)$, $(0~0~\ell)$, and $(h~h~2-h)$ at $T=0.05$~K. The blue area stands for the background. Dashed red and blue plain lines are the result of the fit considering a dual response consisting in quasi-elastic and inelastic contributions. Each contribution is represented in red while the total is represented in blue.}
\label{fig1:inelastic}
\end{figure}

Series of inelastic data were then collected along several high symmetry directions $(h~h~2)$, $(0~0 ~\ell)$, $(1~1~\ell)$, $(h~h~2-h)$ and $(h~h~3-h)$ at the base temperature of $T=0.05$~K. For all measured wave-vectors $\mathbf{Q}$ (see Figure \ref{fig1:elastic}(a), black dots), the spectra show above the elastic line a very broad signal extending up to about $1-1.5$~meV (see Figure \ref{fig1:inelastic}). Despite the rather good energy resolution $\Delta_r$=85 $\mu$eV, {\it no well-defined collective excitations} could be observed. This is in sharp contrast with the resolution-limited spin-waves observed in the field polarized phase of \ybti\, \cite{ross1,rossprx}.

Two qualitatively different behaviors can be observed depending on the wave-vector position relative to the rods of scattering. The spectra taken at $\mathbf{Q}$ vectors located on these rods are dominated by quasi-elastic fluctuations, whereas away from them, the signal appears inelastic. This feature is evidenced on the raw data presented in Figure \ref{fig1:inelastic}, displaying directions $(h~h~2)$, $(h~h~2-h)$ and $(0~0~\ell)$ on the left, middle and right panels respectively. On the left panel, while "flat-toped" at $\mathbf{Q}=(0,0,2)$, the signal becomes quasi-elastic on approaching the rod position at $(2,2,2)$. The same feature is observed for the scans taken along $(h~h~2-h)$ : the line-shape of the spectra becomes quasi-elastic-like while going through the rod position at $\mathbf{Q}=(1,1,1)$ and at $(2,2,0)$. Finally, along $(0,0,\ell)$, i.e. away from the rods, an inelastic line-shape is observed.

A phenomenological fit of the data can be performed considering a dual response consisting in quasi-elastic and inelastic contributions multiplied by the detailed balance factor (see Figure \ref{fig1:inelastic}, plain and dashed lines). Since the experimental resolution is one order of magnitude smaller than the observed excitations, the fitting function was not convolved with the experimental resolution function. The fitting procedure is as follows. First, the width $\Gamma_{QE}\simeq 0.3$~meV of the quasi-elastic response is evaluated using a Lorentzian profile at different wave-vectors along or close to the rods of scattering. $\Gamma_{QE}$ is then considered as wave-vector independent. In addition, a broad Gaussian peak is used to model the inelastic response. The fit is able to converge only if the width of this inelastic peak is fixed to a given value. The free parameters of the fit are thus the intensity of both quasi-elastic and inelastic contributions as well as the peak position. This modeling shows qualitatively the different behaviors close to and away from the rods of scattering, respectively dominated by quasi-elastic and inelastic response. Note that, in such an analysis, the inelastic contribution is peaked at about 0.5~meV and seems not to disperse.

These inelastic data, and especially the flat energy dependence, are for the most part consistent with earlier inelastic results obtained below $\hbar\omega=0.7$~meV around two positions only, $\mathbf{Q} \approx (1.75,1.75,0.5)$ and $(1.5,1.5,1.5)$ (see Figure 4 in Reference \onlinecite{ross2}). The present results demonstrate that the on- and off-rods behaviors can be generalized throughout the Brillouin zone. Furthermore they show that the dynamical response extends up to $1-1.5$~meV (see Appendix \ref{Annexe:neutron} for a further detailed analysis).

\subsubsection{Temperature dependence}

\begin{figure}[!t]
\centerline{
\includegraphics[trim=70px 0 100px 0, width=8cm,clip]{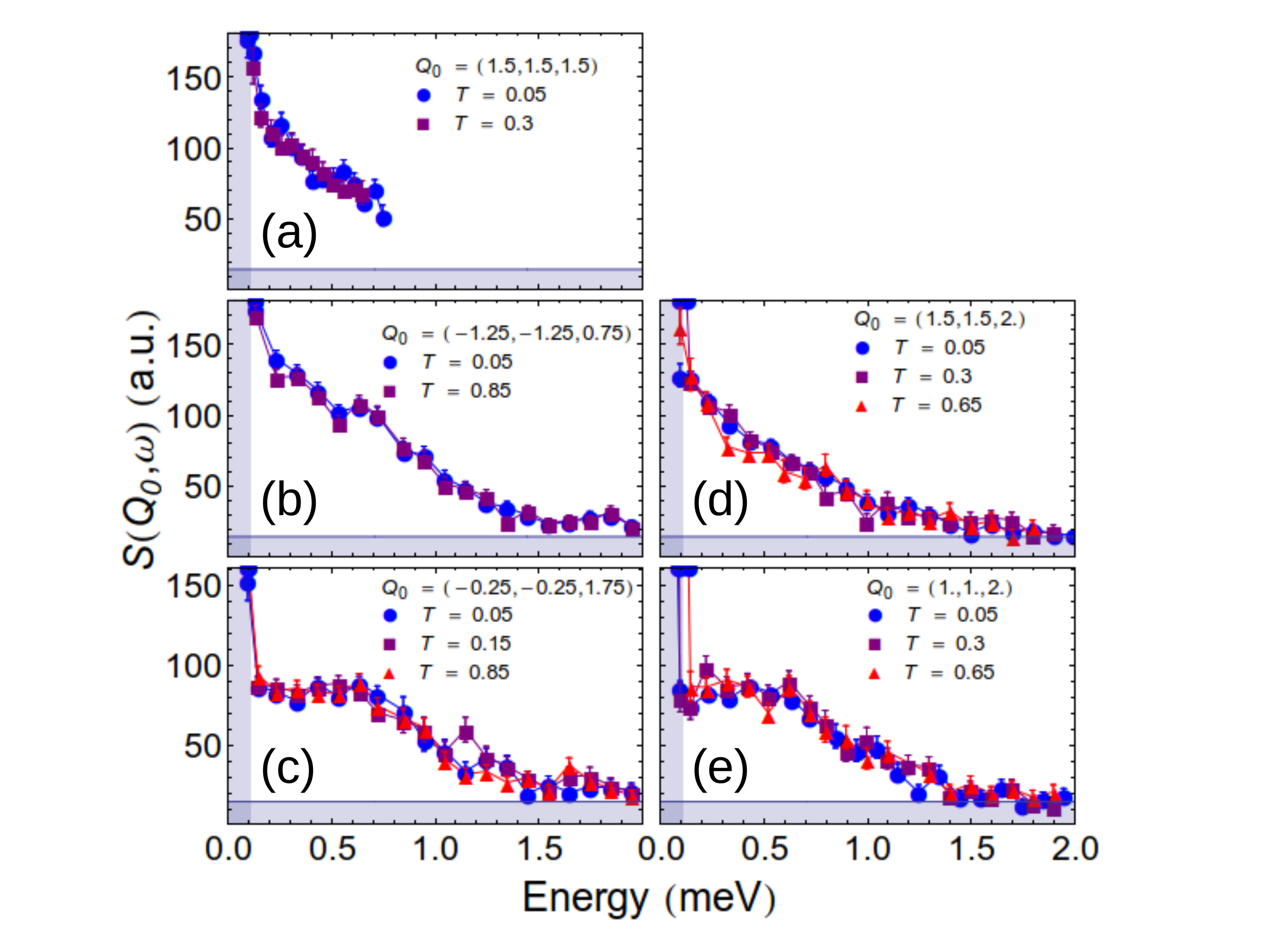}
}
\caption{(Color online) Excitations spectra at several wavevectors close to (a,b,d) or away from (c,e) the rods of diffuse scattering, and temperature from $T=50$ to 850~mK.}
\label{fig5a:TK}
\end{figure}

Interestingly, the spectra recorded at higher temperature do not show any change compared to the low temperature data, indicating that the spin dynamics is only little affected by the increase of temperature up to 850~mK. This remains valid whether $\mathbf{Q}$ lies close to the rods of diffuse scattering or not (see Figure \ref{fig5a:TK} (a,b,d) and (c,e) respectively). 
Although the lack of temperature dependence of the ``off-rod'' scattering is consistent with Ref. \onlinecite{ross2}, the reported ``on-rod'' depletion at energies less than 0.2 meV is not observed in our data. The spectra remain quasi-elastic. As a result we cannot conclude on a slowing down of the spin fluctuations on entering the ordered phase. This difference may be explained by the low transition temperature observed in our sample, as denoted in section \ref{sec:exp_mag}. 

\begin{figure}[!t]
\includegraphics[trim=20px 0 20px 0, clip, width=8.5cm]{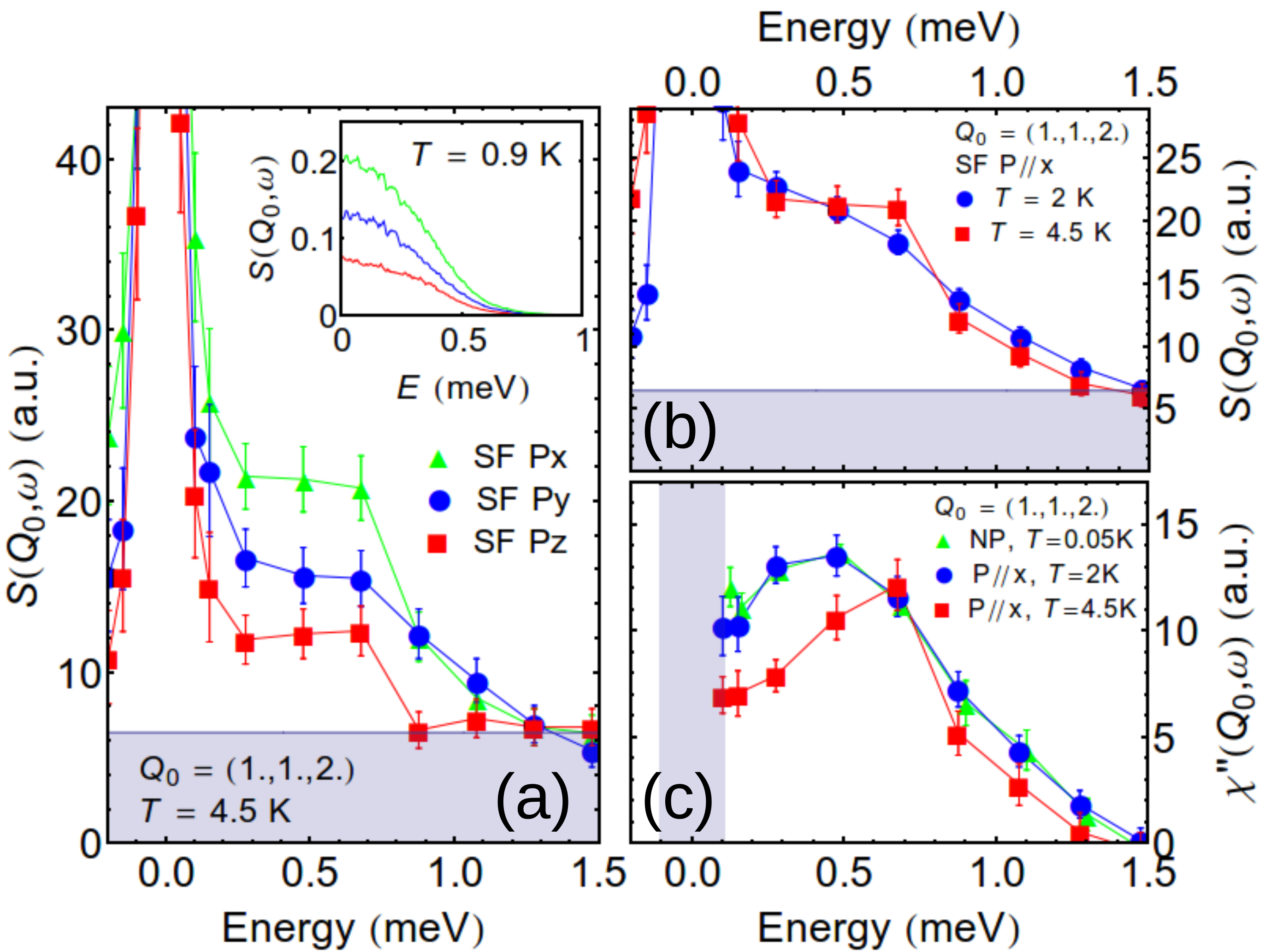}
\caption{(Color online) (a) Spin polarized neutron intensity $S(\mathbf{Q},\omega)$ taken at $\mathbf{Q}=(1,1,2)$ and $T=4.5$~K. Red squares, blue circles and green triangles respectively denote the spin flip contribution in the $z$, $y$ and $x$ polarization channels. The inset shows the calculated spectra, see section \ref{ssec:simus} at $T=0.9$~K. (b) Spin flip intensity with polarization along $x$ at 4.5 (red squares) and 2~K (blue circles). (c) shows the imaginary part of the magnetic susceptibility $\chi''(\mathbf{Q},\omega)$ at $\mathbf{Q}=(1,1,2)$ for 4.5 (red squares), 2 (blue circles) and 0.05~K (green triangles).}
\label{fig5b:TK}
\end{figure}

Inelastic polarized neutron measurements have also been performed at higher temperature ($T=2$ and 4.5~K) at $\mathbf{Q}=(1,1,2)$. The data, shown in Figure \ref{fig5b:TK}, first confirm the magnetic nature of the inelastic signal. At $T=4.5$~K, the full polarization analysis points out that the $S(\mathbf{Q},\omega)_y$ contribution is a little bit more intense than the $S(\mathbf{Q},\omega)_z$. Surprisingly, it also appears that $S(\mathbf{Q},\omega)_z$ (red curve) vanishes above $\hbar \omega_{max} \simeq 0.75$~meV while $S(\mathbf{Q},\omega)_y$ (blue curve) extends up to $\simeq 1.1$~meV. A depletion of the spectrum is also observed at low energies, making the generalized susceptibility $\chi''(\mathbf{Q},\omega)=S(\mathbf{Q},\omega)/\left(1+n(\omega)\right)$ quite narrow, with a maximum around $\hbar\omega=0.6$~meV (see Figure \ref{fig5b:TK} (c)). 
Interestingly, this approximately well defined excitation observed at $T=4.5$~K spreads out and becomes less defined with decreasing temperature (cf. Figure \ref{fig5b:TK}(b)). At $T=0.05$ and 2~K, $\chi''(\mathbf{Q},\omega)$ is superimposed, showing that the spectrum does not evolve anymore below 2~K (see panel (c)). 

\subsubsection{A possible continuum of excitations}

In summary, this study of \ybti\, points out very unconventional spin dynamics. Starting from the high temperature, the spin excitations spectrum broadens below $T_0 \simeq2-4$~K. This is largely unexpected, since, in general, the width of inelastic spectra, related to the rate of the spin fluctuations, tends to increase with increasing temperature. The crossover temperature $T_0$ between the two regimes coincides with the specific heat bump observed around $T=2.5$~K \cite{blote}, below which short range spin ice like correlations establish \cite{applegate} and the rods of diffuse scattering rise \cite{bonville,ross2,thompson}. At low temperature, no well defined excitations are observed but rather a continuum of excitations. Finally, entering the ordered phase does not affect the spectra. 

\section{Numerical support}
\label{sec:num}

These experimental results are now confronted to numerical calculations. First, a model of exchange interactions at play in \ybti\, is proposed. Based on RPA numerical simulations \cite{ersn,erti}, we show that a large range of parameters accounts for the available neutron data obtained in the conventional polarized phase induced by applying magnetic fields of 2 and 5T. Reproducing the strong quasi-elastic scattering around $\mathbf{Q}=(2,2,0)$ with Monte Carlo Spin Dynamics simulations \cite{juju,juju2,tao,conlon,mueller} allows however to constrain further the set of parameters and determine an optimal set of coupling constants. In zero field, these simulations show that \ybti~ is a canted ferromagnet, which yet lies very close to an antiferromagnetic phase. Finally, RPA and classical spin dynamics calculations are presented. Although the latter reproduce some of the experimental features, they both fail to explain the very large energy range of the experimental data. This points out the role of quantum fluctuations that may be amplified by the proximity to a competing antiferromagnetic phase, and which are not captured using a classical approach.
\subsection{Models and Hamiltonians}
\label{ssec:model}

The starting point is the widely accepted Hamiltonian for pyrochlore systems \cite{gingrasrmp}:
\begin{eqnarray}
H & = & H_{CEF} + H_{exc} + H_Z.
\label{gen_ham}
\end{eqnarray}
$H_Z= g_J \sum_i \bm H \cdot \bm J_i$, is the Zeeman term, with $\bm H$ the applied magnetic field and $\bm J_i$ the magnetic moment at site $i$. $H_{CEF}=\sum_i \sum_{k,q=1}^6 B_{k}^q \mathcal{O}_{k,i}^q$ is the crystal electric field (CEF) Hamiltonian, with $\mathcal{O}_{k}^q$ the Stevens operators and $B_{k}^q$ scalar parameters \cite{wybourne}. Several estimations of the $B_k^q$ parameters may be found in the literature \cite{cao1,hodges01,malkin10,bertin}. In the present paper, we use the CEF parameters $(B_2^0, B_4^0, B_4^3, B_6^0, B_6^3, B_6^6)=(536, 7752, -2942, 830, 671, 739)$~K obtained in ref. \onlinecite{bertin}. These parameters lead to a doublet CEF ground-state, separated from the first excited level by an energy gap $\Delta E \simeq 700$~K. The magneto-crystalline anisotropy of the ground-state doublet is XXZ-like, with an easy plane perpendicular to the local $\langle 111 \rangle$ directions. The $g_{z}/g_{\perp}\simeq 0.504$ ratio denotes a weak in-plane anisotropy with Land\'e factors $g_{z}=2.06$ and $g_{\perp}=4.09$.
$H_{exc} = \sum_{ij} \bm J_i \tilde{\mathcal{K}}_{ij} \bm J_j$ is a bilinear coupling Hamiltonian, where the interaction tensor $\tilde{\mathcal{K}}_{ij}$ couples next neighbors magnetic moments $\bm J$ at sites $i$ and $j$. Note that $H_{exc}$ gathers all possible physical (e.g. exchange and dipolar coupling restricted to nearest neighbors), whether they are isotropic or not. By symmetry arguments, the 9 coupling constants of the $3 \times 3$ tensor $\tilde{\mathcal{K}}_{ij}$ are reduced to only four\cite{curnoe}. Here, we assume an exchange tensor which is diagonal in the ($\mathbf{a},\mathbf{b},\mathbf{c})$ frame linked with a R-R bond \cite{malkin04}:
\begin{eqnarray*}
\mathbf{J}_i {\cal \tilde K} \mathbf{J}_j &=& \sum_{\mu,\nu=x,y,z} J_i^{\mu} 
\left( 
{\cal K}_a a_{ij}^{\mu} a_{ij}^{\nu} + 
{\cal K}_b b_{ij}^{\mu} b_{ij}^{\nu} + {\cal K}_c c_{ij}^{\mu} c_{ij}^{\nu} 
\right) J_j^{\nu} \\
& &- {\cal K}_4 \sqrt{2}~\vec{b}_{ij}.(\vec{J}_i \times \vec{J}_j)
\end{eqnarray*}

Since the energy gap between the CEF ground-state and the first excited levels are order of magnitudes larger than the exchange interactions and the Zeeman term, it is possible to define effective spin $1/2$ operators, denoted by $\mathbf{S}_i$, by projecting the full moment $\mathbf{J}_i$ onto the CEF ground-state doublet. As a result, an effective Hamiltonian can be defined in terms of new anisotropic couplings between the $\mathbf{S}_i$ spin components:
\begin{eqnarray}
H_{\mbox{eff}} & = & \sum_{ij} \mathbf{S}_i \tilde{J_{ij}} \mathbf{S}_j,
\label{eq:effective}
\end{eqnarray}
A popular convention used in Ref \onlinecite{curnoe,rossprx,savary,wong} consists in using $({\sf J}_{\pm\pm},{\sf J}_{\pm},{\sf J}_{z\pm},{\sf J}_{zz})$ defined as:
\begin{eqnarray*}
H_{\mbox{eff}} &=& \sum_{i,j} {\sf J}_{zz} {\sf S}^z_i {\sf S}^z_j - {\sf J}_{\pm} \left({\sf S}^+_i {\sf S}^-_j + {\sf S}^-_i {\sf S}^+_j \right) \\
& &
+ {\sf J}_{\pm\pm} \left(\gamma_{ij} {\sf S}^+_i {\sf S}^+_j + \gamma^*_{ij} {\sf S}^-_i {\sf S}^-_j \right) \\
& &
+ {\sf J}_{z \pm} \left[ {\sf S}_i^z \left( \zeta_{ij} {\sf S}^+_j + \zeta^*_{ij} {\sf S}^-_j\right) + i \leftrightarrow j \right] 
\end{eqnarray*} 
where $\gamma_{ij}, \zeta_{ij}$ are c-numbers \cite{rossprx,savary,wong}. Note that "sanserif" notations refer to local bases. $({\sf J}_{\pm\pm},{\sf J}_{\pm},{\sf J}_{z\pm},{\sf J}_{zz})$ are related to ${\cal K}_{a,b,c,4}$ by the following relations \cite{ersn}:
\begin{eqnarray*}
{\sf J}_{zz}  & = & \lambda_z^2 ~\frac{{\cal K}_a-2{\cal K}_c-4{\cal K}_4}{3} \\
{\sf J}_{\pm} & = & -\lambda_{\perp}^2 ~\frac{2{\cal K}_a-3{\cal K}_b-{\cal K}_c+4{\cal K}_4}{12} \\
{\sf J}_{z\pm} & = & \lambda_{\perp}~\lambda_z ~\frac{{\cal K}_a+{\cal K}_c-{\cal K}_4}{3 \sqrt{2}} \\
{\sf J}_{\pm \pm} & = & \lambda_{\perp}^2 ~\frac{2{\cal K}_a+3{\cal K}_b-{\cal K}_c+4{\cal K}_4}{12}
\end{eqnarray*}
with $\lambda_{\perp,z}=\frac{g_{\perp,z}}{g_J}$.

In the following, we will follow the alternative convention of Ref \onlinecite{rossprx,ludo}, using the set of effective parameters $J_{1,2,3,4}$ (correspondence between the different conventions are given in Appendix \ref{Annexe:conversion}): 
\begin{eqnarray*}
J_{1} & = & \frac{1}{3} \left( 2{\sf J}_{\pm \pm} + 4 {\sf J}_{\pm} + 2\sqrt{2} {\sf J}_{z\pm} - {\sf J}_{zz} \right) \\
J_{2} & = & \frac{1}{3} \left( 4{\sf J}_{\pm \pm} - 4 {\sf J}_{\pm} + 4\sqrt{2} {\sf J}_{z\pm} + {\sf J}_{zz} \right) \\
J_{3} & = & \frac{1}{3} \left(-4{\sf J}_{\pm \pm} - 2 {\sf J}_{\pm} + 2\sqrt{2} {\sf J}_{z\pm} - {\sf J}_{zz} \right) \\
J_{4} & = & \frac{1}{3} \left(2{\sf J}_{\pm \pm} - 2 {\sf J}_{\pm} - 2\sqrt{2} {\sf J}_{z\pm} - {\sf J}_{zz} \right)
\end{eqnarray*}

\begin{figure*}
\includegraphics[keepaspectratio=1, width=\textwidth]{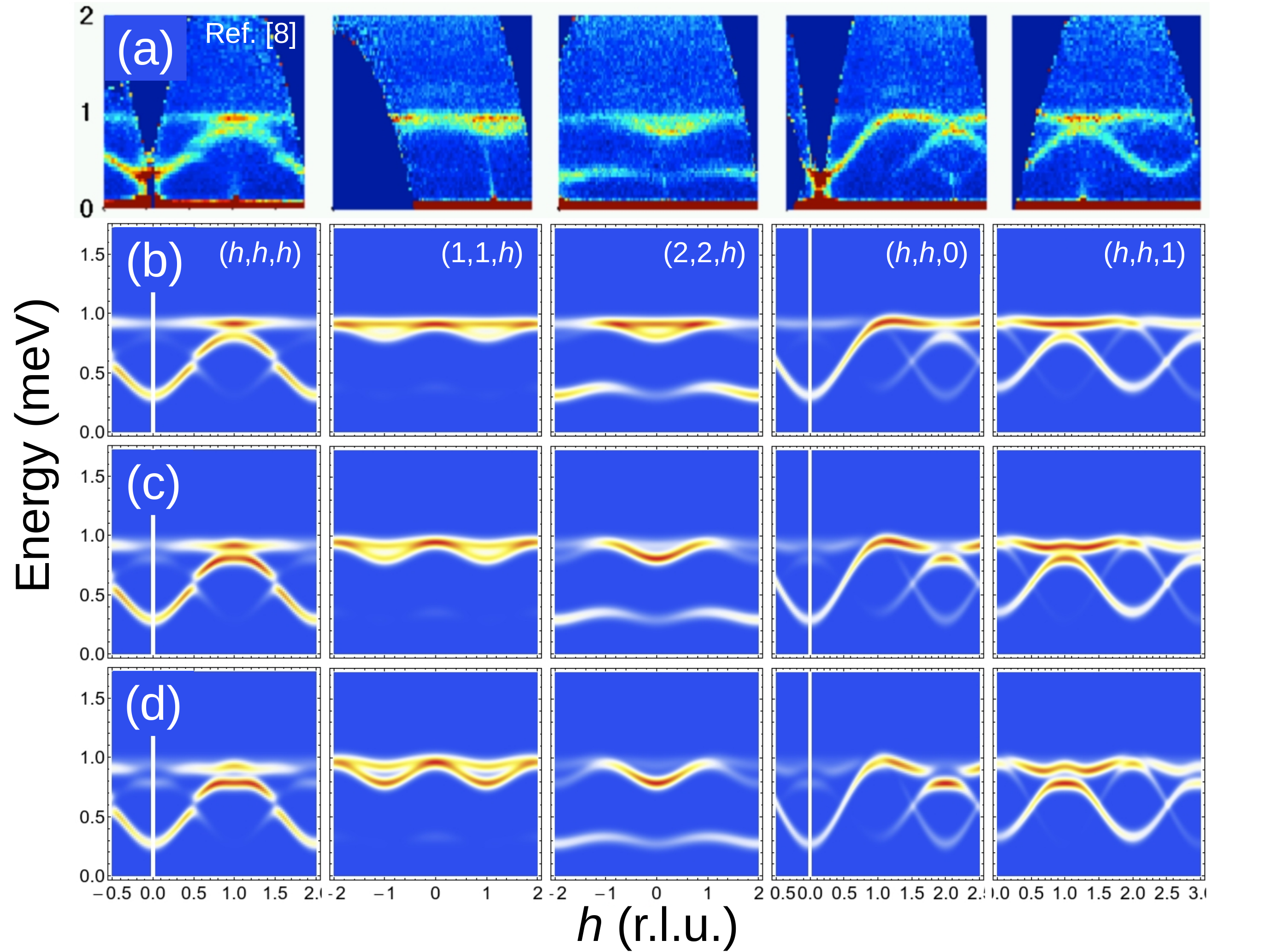}
\caption{(Color online) (a) Experimental spectra obtained in reference \onlinecite{rossprx} at $H=2$~T and $T=0.05$~mK. (b,c,d) Examples of simulations, performed in the same temperature and field conditions, giving a good agreement with the data : 
(b)  $(J_1, J_2, J_3, J_4)= (-0.09, -0.19, -0.25, 0.005
)$; (c), $(J_1, J_2, J_3, J_4)=(0.04, -0.29, -0.26, 0.024)$, and (d): $(J_1, J_2, J_3, J_4)=(-0.02,-0.34,-0.29, 0.036)$~meV.
}
\label{fig:SWField}
\end{figure*}


\subsection{Overview of the phase diagram}

The zero-temperature phase diagram for the nearest-neighbor pseudo spin-1/2 Hamiltonian given by Eq. (\ref{eq:effective}) has been studied in Ref. \onlinecite{wong,ludo}. Owing to the XXZ nature of the \yb~ magnetic anisotropy, and depending on the sign and amplitude of the exchange parameters, four different magnetic phases are obtained in the phase diagram: a canted (or splayed) ferromagnetic phase (labeled FM), the antiferromagnetic Palmer-Chalker phase which results from the so-called $\Gamma_7$ representation \cite{pc}, and the basis states $\psi_2$ and $\psi_3$ of the $\Gamma_5$ representation \cite{champion}, whose combination form the classically continuously degenerated manifold. Figure \ref{fig7:phasediag} shows a sketch of the $\psi_2$, $\psi_3$ and FM spin configurations on one tetrahedron.

As quoted previously, the parameters of the Hamiltonian determined for \ybti~ in prior studies place it in the ferromagnetic phase. The aim of the next paragraphs is to test the unicity of this set of parameters, and further explore the consequences of the location of \ybti~ in the phase diagram.

\subsection{Exchange couplings in \ybti\, }
\label{ssec:parameters}

As already reported in Ref. \onlinecite{ross1,rossprx}, a polarized state sets in when a magnetic field $H > H^c_{T=0.05K}=0.5$~T is applied along $[1~\bar 1~ 0]$. The spin dynamics consists in spin-wave excitations propagating in different directions of reciprocal space, as measured by inelastic neutron scattering for $H=2$ and 5~T \cite{rossprx}. The data collected in these studies along directions $(h~h~h)$, $(1~1~\ell)$, $(2~2~\ell)$, $(h~h~0)$ and $(h~h~1)$ are shown in Figure \ref{fig:SWField}(a). Four modes can be distinguished in the experimental spectra, one of them being almost flat throughout the Brillouin Zone into the $(h~h~0)-(0~0~\ell)$ plane.

To determine the exchange couplings, we solve the general Hamiltonian (\ref{gen_ham}) in the RPA approximation \cite{ersn,erti}, and perform exhaustive calculations of the spin-spin correlation function $S(\bm Q,\omega)$. The quality of the fit relies on a good matching of the location of the maximum neutron intensity in $(h~h~h)$, $(1~1~\ell)$, $(2~2~\ell)$, $(h~h~0)$ and $(h~h~1)$ directions, as well as of the overall spectral weight \cite{erti}. The calculated mean-field magnetic moment is checked to reproduce the experimental one \cite{elsa} at $T=0.05$~K for $H=2$ and 5~T. Finally ${\cal K}_{a,b,c,4}$ are transformed into $J_{1,2,3,4}$.

\subsubsection*{Reduction of the parameter space by fitting the field-induced conventional spin-wave excitations}
In contrast with prior reports, we find that this procedure does not lead to a unique solution. Representative spectra are shown in Figure \ref{fig:SWField} (b,c,d) for different relevant sets. 
They all reproduce most of the experimental features, yet slight differences, mainly in the spectral weight distribution remain. For instance, while the spectral weight of the two higher branches in the $(1~1~\ell)$ direction is minimum at $\mathbf{Q}=(1,1,0)$, the inverse is observed in the simulations. More generally, the spectral weight distribution of the modes is never fully reproduced, although the agreement seems slightly better in the intermediate parameter range (e.g. Figure \ref{fig:SWField} (c)). 

\begin{figure}
\centerline{
\includegraphics[keepaspectratio=1, width=9.5cm]{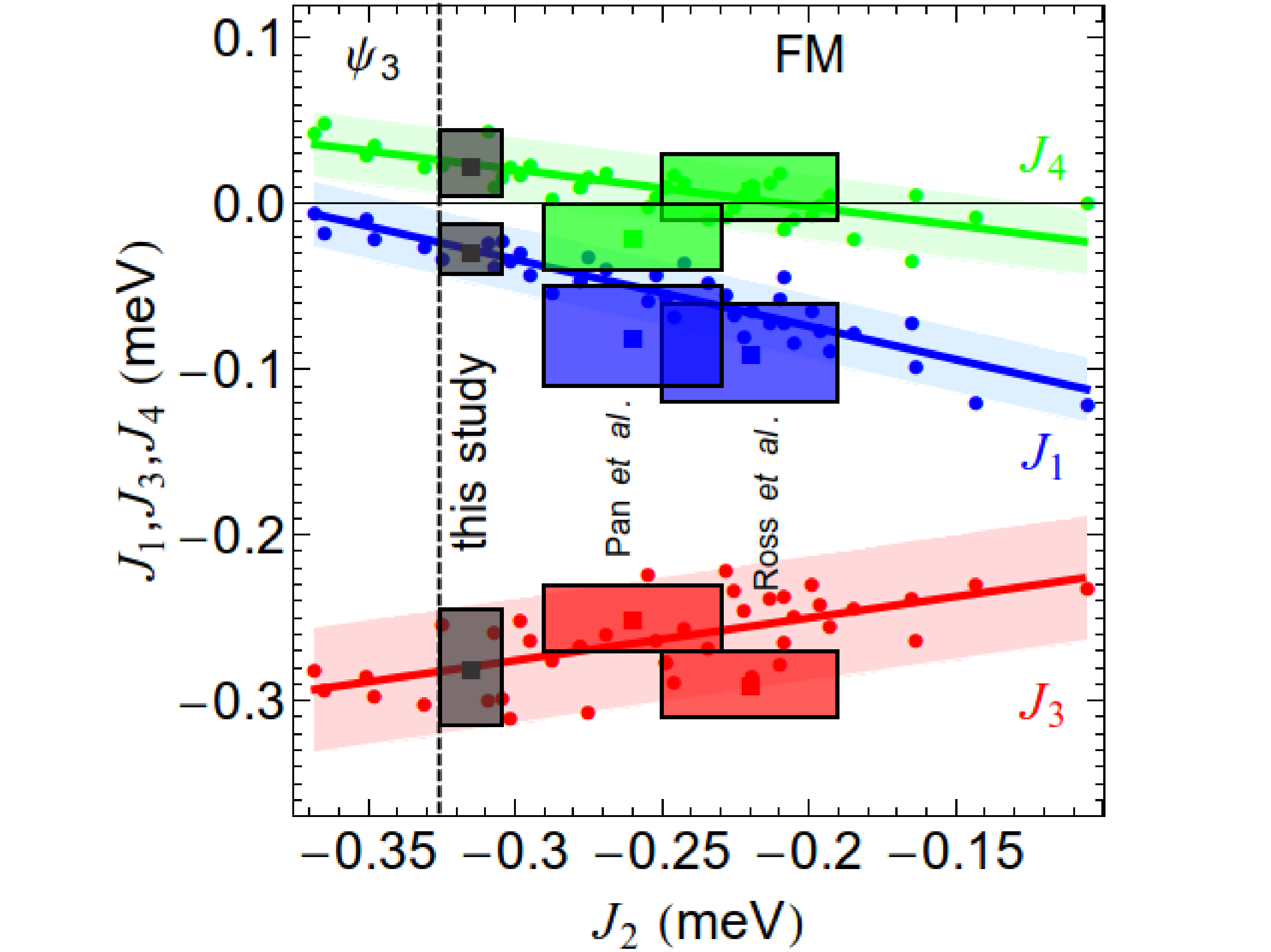}
}
\caption{(Color online) Exchange interaction sets determined based on the field-induced spin-waves at $H=2$ and 5~T. Exchange interactions reported in Ref. \onlinecite{rossprx,pan} are indicated. Those reported in Ref. \onlinecite{thompson,chang} are out of the parameter range of the figure, and are instead indicated in the Fig. \ref{corres} of Appendix \ref{Annexe:conversion}.}
\label{fig:sets}
\end{figure}

The sets of parameters giving such good agreement are reported in Figure \ref{fig:sets}. 
Clearly, there is a linear relation between the $J_i$'s, so that the 4 parameters space is now reduced to a one dimensional space. Colored areas delimit the uncertainties. Note also that, since both magnetic fields stabilize the same ferromagnetic ground-state, fitting the $H=2$ and 5~T data provides similar constraints in the parameter space, resulting in the same linear relation between coupling constants. For this reason, the parameters obtained from the $H=2$ and $5$~T spectra are not distinguished and represented together in Figure \ref{fig:sets}. 
It must be emphasized that the set of parameters reported in Ref. \onlinecite{rossprx} is consistent with our determination. 

It should be stressed also that a number of parameters do not lead, in zero field, to the expected ferromagnetic phase (labeled ``FM'' in Figure \ref{fig:sets}) but to the $\psi_3$ antiferromagnetic ground-state. The phase boundary separating the $\psi_3$ and the FM phases at $T=0$, while exploring the one dimensional parameter space described above, is depicted in Figure \ref{fig:sets} by a vertical dashed line and corresponds to ${J_i}^c = (-0.023,-0.326,-0.282,0.026)$~meV.

\begin{figure}
\centerline{
\includegraphics[trim=90px 0 120px 0, clip, width=8.5cm]{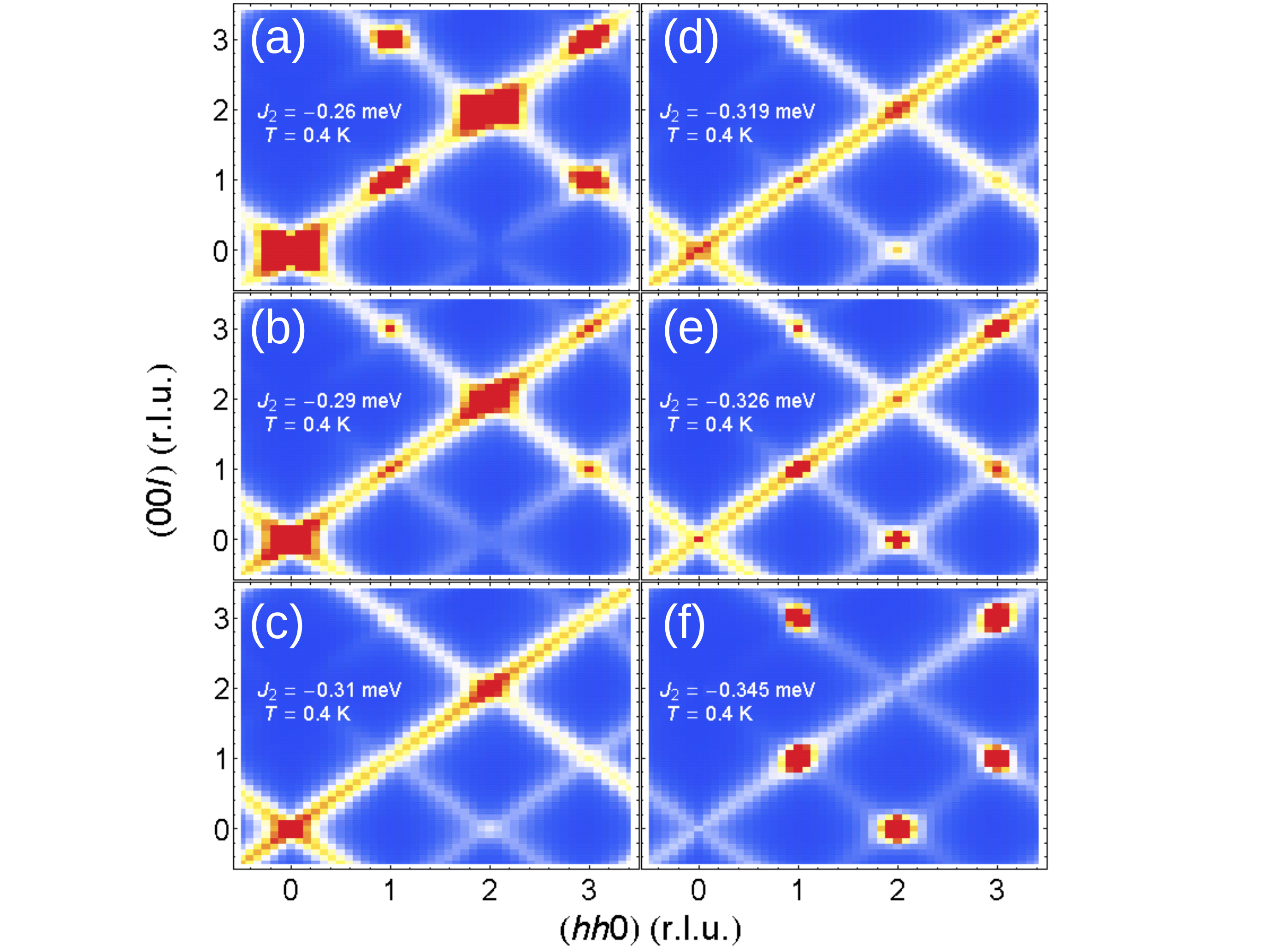}
}
\caption{(Color online) Elastic scattering function $S(\bm Q,\omega=0)$ in the $(h,h,l)$ plane obtained from spin dynamics simulations for $J_2=-0.26, -0.29, -0.31, -0.319, -0.326, -0.345$~meV at $T=0.4$~K.}
\label{fig:diffus}
\end{figure}

\subsubsection*{Rods of diffuse scattering as a further constraint on the coupling parameters}

To further constrain the dimension of parameter space, other experimental results have to be taken into account. To this end, $S(\mathbf{Q},\omega=0)$ was calculated using a combination of Monte Carlo and spin dynamics simulations at finite temperature, aiming at reproducing the rods of diffuse scattering. This numerical method is briefly described in Appendix \ref{Annexe:SpinDynamics} and more details may be found in Ref. \onlinecite{juju,juju2}. Calculations are based on the effective Hamiltonian (\ref{eq:effective}) and were performed in the short-range correlated regime at $T=0.4$~K, just above the critical temperature, for different coupling sets within the one dimensional parameter space determined above.
Those numerical simulations (see Figure \ref{fig:diffus}) show that the rods of scattering around $(2,2,2)$ and $(2,2,0)$ are simultaneously observed when approaching the FM/AFM phase boundary (in agreement with Ref. \onlinecite{ludo}). This confirms the observation made in section \ref{sec:exp_ela} of coexisting FM and AFM short range correlations. The cuts in direction $(h~h~2-h)$ shown in Figure \ref{fig1:elastic} (b) and (c) further show that a good qualitative agreement between the experimental data (filled circles) and the spin dynamics simulations (empty circles) may only be achieved while lying very close to the FM/AFM phase boundary.

Combining the different results, the best agreement is obtained for ${J_i}^0 = (-0.03(2), -0.32(1), -0.28(3), 0.02(2))$~meV (grey area in Figure \ref{fig:sets} (d)). The associated Curie-Weiss temperature is $\theta_{CW}=0.53(1)$~K, in excellent agreement with the experimental value $\theta_{CW}\simeq 0.5$~K determined in section \ref{sec:exp_mag} (see also Ref. therein). It is worth noting that this set of values locates \ybti~ closer to the AFM/FM phase boundary than the one obtained in Ref. \onlinecite{rossprx,pan}. It is also quite different from that obtained by Thompson {\it et al} \cite{thompson} as well as that reported by Chang {\it et al} \cite{chang}. Actually, the latter studies yield spin wave spectra in strong disagreement with experiment, while our set reconciles both the diffuse scattering and inelastic neutron data.

\begin{figure}
\centerline{
\includegraphics[trim=100px 0 130px 0px, clip, width=8.5cm]{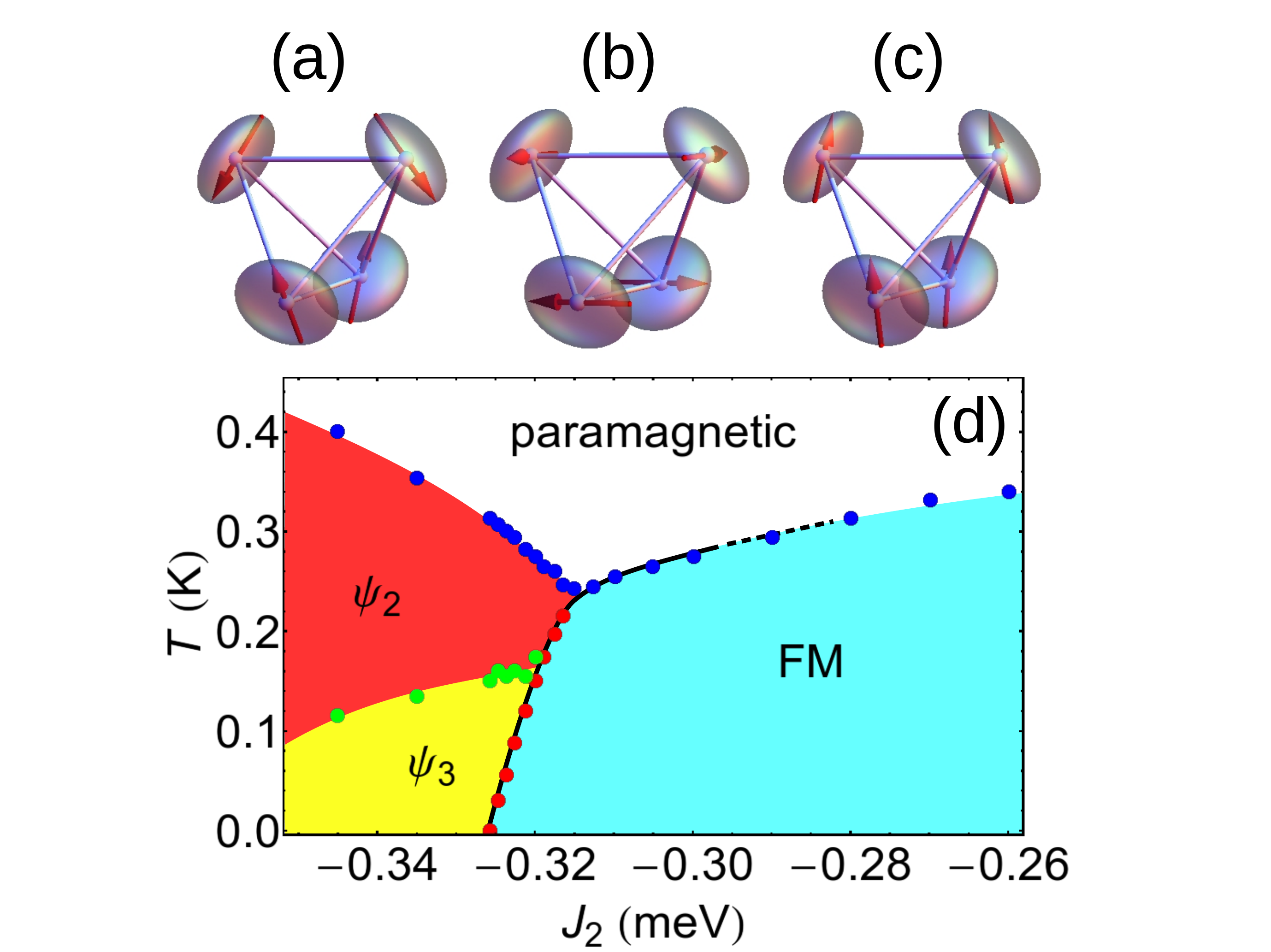}
}
\caption{
(Color online) Sketches of the spin configurations on one tetrahedron in the (a) $\psi_2$, (b) $\psi_3$ and (c) FM phases. (d) Phase diagram calculated as a function of temperature and of exchange coupling varying along the optimal one dimensional parameter space determined from the spin-wave fitting procedure. The black line roughly indicates the range of parameters for which the transition toward the ferromagnetic order is first order.}
\label{fig7:phasediag}
\end{figure}
\subsection{Zero field Ground-state. Order of the transition}

We now examine in more detail the consequences of the above results on the zero field ground state. To this end, the phase diagram within the 1D parameter space discussed in the previous section is calculated based on Monte Carlo simulations, for different lattice size up to $L=16$, $N_{sp}=65536$ (see Figure \ref{fig7:phasediag}). Since the critical temperatures does not seem to evolve much with the lattice size for $L>12$ (not shown), the phase boundaries have been roughly determined by locating the maximum of the specific heat for $L=12$\cite{footnote}.
For the optimal values ${J_i}^0$ close to the boundary located at ${J_i}^c$, the specific heat and magnetic susceptibility show, upon cooling, a succession of several transitions. Two second order transitions first occurs at about $T=0.275(3)$ and $0.174(3)$~K towards the $\psi_2$ and $\psi_3$ AFM states respectively. This is followed by an abrupt transition at $T_c=0.15(1)$~K towards a FM state (sketches of the magnetic configuration are shown in Figure \ref{fig7:phasediag} (a,b,c)). The first order character of this second transition is demonstrated by a finite latent energy at the critical temperature. The latent energy decreases while getting deeper into the FM state, such that the first order transition smoothly transforms into a second order one. This behavior while approaching the phase boundary is consistent with the study by Yan {et al.} \cite{ludo}.

Both the reentrant behavior of the FM phase at low temperatures and the evolution of the order of the transition on approaching the phase boundary might result from thermal order by disorder: although the ferromagnetic state has a lower internal energy for $J_2 > J_2^c$, the $\psi_2/\psi_3$ phases may be favored in a small temperature range because of their strong antiferromagnetic fluctuations; the long-range ferromagnetic order is then stabilized at lower temperatures where thermal fluctuations are reduced.

%



\subsection{Spin dynamics simulations in zero field}
\label{ssec:simus}


We finally turn to the calculations of the classical spin dynamics aiming at a final discussion of the experimental results. 

\subsubsection*{RPA approximation}

\begin{figure}
\includegraphics[trim=70px 0 80px 0, clip, width=8.5cm]{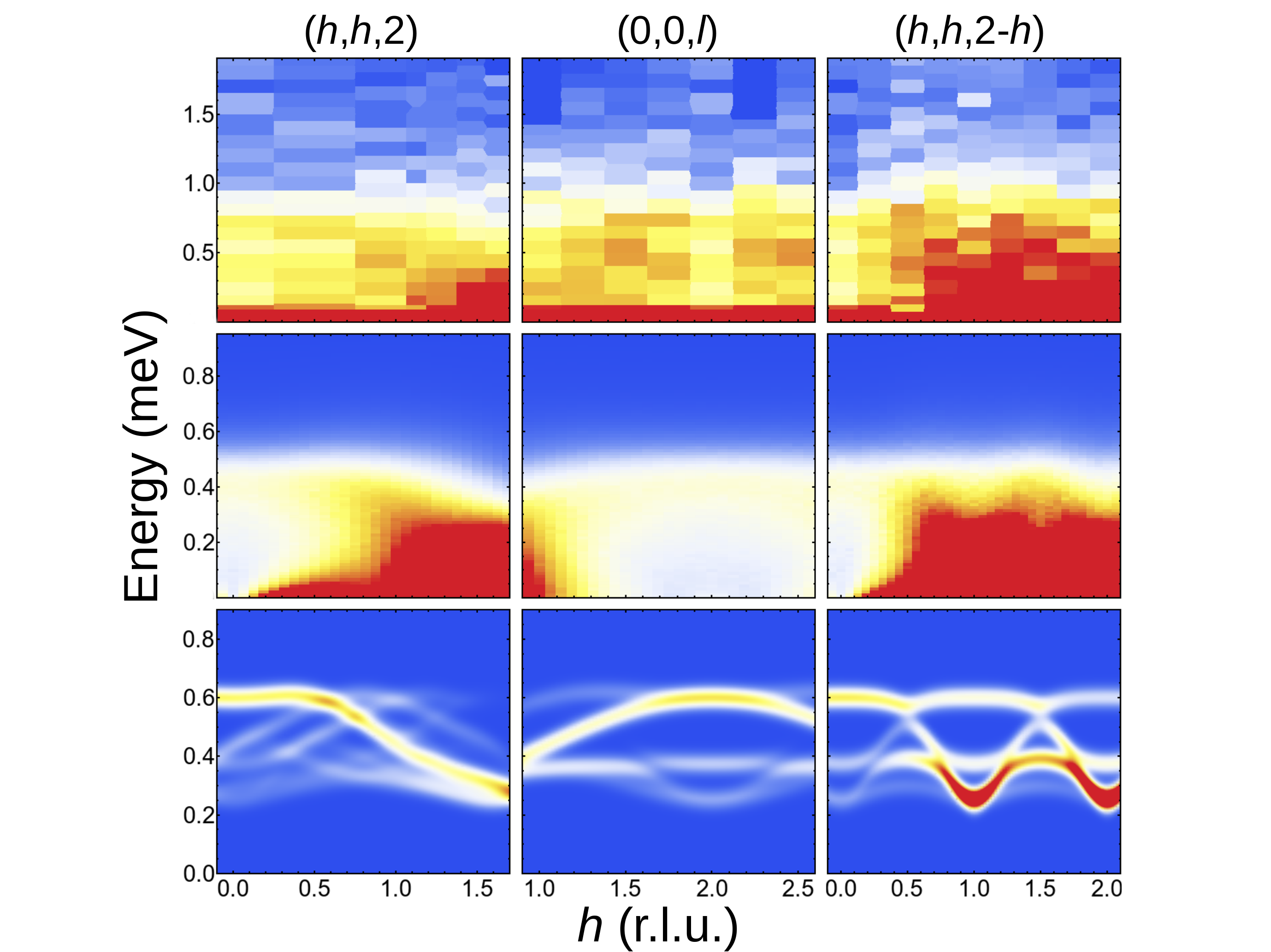}
\caption{(Color online) $S(\mathbf{Q},\omega)$ along high symmetry directions : (a) neutron data (this work) taken at $T=0.05$~K; (b) calculated spectra obtained from classical spin dynamics simulations in the short-range correlated regime ($T=0.4$~K); (c) RPA spectra obtained in the FM phase.}
\label{fig10:SDMaps}
\end{figure}

$S(\mathbf{Q},\omega)$ was first calculated in the RPA approximation. Spectra have been averaged over equi-populated domains. They are presented in Figure \ref{fig10:SDMaps}(c) for different high-symmetry directions in reciprocal space. For comparison, experimental data in the same directions are shown in panel (a).
Calculations have also been performed for different exchange interactions around the FM/AFM phase boundary, along the 1D-parameter space determined in the previous section. While the excitations are ``gapless'' in the AFM phase (we did not consider order by disorder or crystal field effects at this point, which could however be at the origin of a small spin gap), an anisotropy gap opens in the FM region because of a sizeable tilt of the spins from their easy-plane (see Figure \ref{fig10:SDMaps}(c)).
Obviously, at this level of approximation, the model fails to reproduce the experimental data. Actually, the calculated spectra consist in conventional spin wave excitations, which cannot capture the flat line shape of the experimental spectra. A more striking discrepancy especially affects the energy scale, which is about twice larger experimentally than in the simulations at most wave-vectors. In the simulations, the highest spin-wave branch is observed at $\hbar\omega=0.6$~meV, a value which does not evolve much by varying the exchange parameters. Experimentally, the spectral weight reaches energies $\hbar\omega \simeq 1-1.5$~meV depending on the wave-vector (see Figure \ref{fig1:inelastic}). Note that such a difference cannot be explained by the experimental resolution, which is around $\Delta E\simeq 0.095$~meV for $k_f=1.15$~\AA~ and an energy transfer $\hbar\omega=1$~meV.

\subsubsection*{Finite temperature spin dynamics simulations in the short-range correlated regime}

\begin{figure}
\includegraphics[keepaspectratio=1, width=8.5cm]{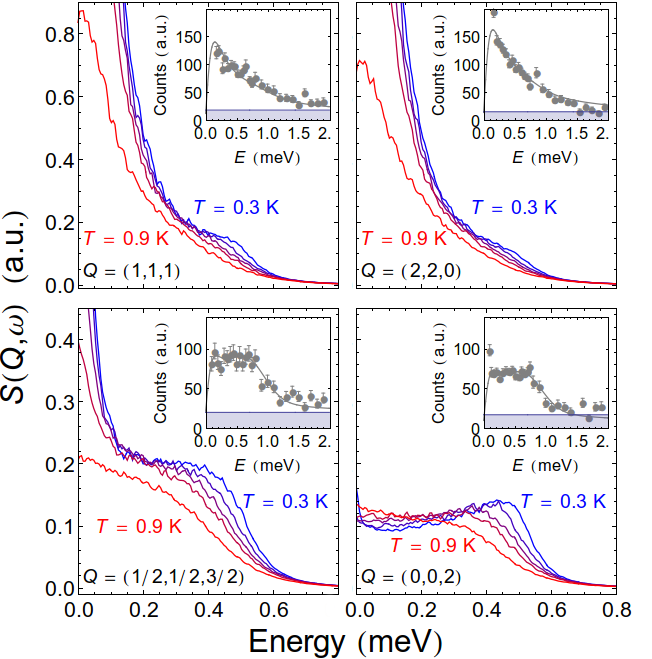}
\caption{(Color online) Calculated $S(\mathbf{Q},\omega)$ obtained from classical spin dynamics simulations at different temperatures $T=0.3,0.4,0.5,0.6$ and $0.9$~K, and for relevant wave-vectors $\mathbf{Q}=(1,1,1)$, $(2,2,0)$, $(1/2,1/2,3/2)$, and $(0,0,2)$. Insets show the neutron data at the same wave-vectors.}
\label{fig11:SDvsT}
\end{figure}

To go a step further, the classical spin dynamics was solved numerically by taking into account non-linear effects associated with thermal fluctuations (see Appendix \ref{Annexe:SpinDynamics} and Ref. \onlinecite{juju,juju2} for technical details about the method). Such spin dynamics simulations of $S(\mathbf{Q},\omega)$ have been performed at different temperatures from $T=0.1$ to $0.9$~K, on the basis of the effective Hamiltonian (\ref{eq:effective}) and optimal exchange parameters ${J_i}^0$ proposed in the previous section.
Intensity maps of $S(\mathbf{Q},\omega)$ are shown Figure \ref{fig10:SDMaps} (b) for $T=0.4$~K in high symmetry directions. The simulations succeed in reproducing the overall shape of the excitation spectrum observed experimentally: $(i)$ in direction $(h~h~2)$, the excitation spectrum goes from quasi-elastic approaching $\mathbf{Q}=(2,2,2)$ to flat-toped excitations at $(0,0,2)$; $(ii)$ the excitations are mainly flat or weakly inelastic along direction $(0~0~\ell)$; $(iii)$ in direction $(h~h~2-h)$, a strong quasi-elastic signal overwhelm the spectra between $\mathbf{Q}=(1,1,1)$ and $(2,2,0)$. 
More generally, the computed $S(\mathbf{Q},\omega)$ displays a quasi-elastic (resp. flat-toped/weakly inelastic) line-shape close to (resp. away from) the rods of scattering, as observed experimentally. 
This is illustrated Figure \ref{fig11:SDvsT}, displaying the calculated scattering function for different temperatures $T=0.3,0.4,0.5,0.6$ and $0.9$~K, and wave-vectors $\mathbf{Q}=(1,1,1)$, $(2,2,0)$, $(1/2,1/2,3/2)$, and $(0,0,2)$. While the two first positions are dominated by a strong quasi-elastic contribution, the latter are mostly inelastic. The experimental data measured at the same wave-vectors are shown in the insets.
Moreover, the calculated spectra do not evolve much with temperature below $T=0.6$~K. Experimentally, this behavior is even more impressive since the spectra do not depend on temperature between $T=0.05$ and 2~K.

However, despite this success, the spin dynamics simulations fail to reproduce the large energy range, which does not exceed $\hbar\omega\simeq0.6$~meV in the calculations, as also observed in RPA simulations. 

\section{Discussion}

In summary, our results show that for the optimal values of coupling constants determined to reproduce the field induced spin wave spectra, classical calculations captures some of the magnetic properties of \ybti. It reproduces different experimental features, such as the non-trivial structure of the diffuse scattering which points out coexisting FM and AFM correlations above $T_c$, as well as the transition towards a long-range ordered canted ferromagnetic state at $T_c$.
At lower temperature, thermal AFM fluctuations have then been shown to affect the stability of the FM order: approaching the boundary from the FM side, the order of the transition smoothly goes from second to first order; sufficiently close to the boundary, a long ranged antiferromagnetic order may even be favored in a significant temperature range above $T_c$. 
This succession of phase transitions observed close to the phase boundary could clarify the peculiar shape of the magnetization curve as a function of temperature on a single crystal recently reported (see Figure 2(a) of Ref. \onlinecite{elsa}): with decreasing temperature, it shows first a reversible bump around $T=0.18$~K, followed by the irreversible step at $T_c=0.16$~K associated with the long range ferromagnetic order.

However, the classical approach fails to explain the strongly reduced ordered moment observed experimentally below $T_c$ 
as well as the unconventional nature of the spin dynamics, not only at very low temperatures, but up to $T_0\simeq2$~K. Both features are obviously connected, since the part of the moment which is not condensed in the static component remains, by essence, fluctuating. 

To explain this unconventional spin dynamics, the coupling to other degrees of freedom cannot be ruled out. This phenomena has been shown to be of utmost importance in \tbti\, for instance \cite{guitteny,Fennell3}, but there is, to our knowledge, no experimental evidence in favor of such a coupling in \ybti. Moreover, the unconventional dynamics appears around 2~K which coincides with the apparition of short range correlations, which is a clue for a purely magnetic origin.

It is thus tempting to invoke quantum effects as the missing ingredient in the proposed model. Indeed, the proximity with the AFM magnetic states is likely a source of enhanced quantum fluctuations. To account for the unconventional spin dynamics, two scenarii come to mind: $(i)$ the fractionalization of magnons into deconfined new spin 1/2 particles and $(ii$) the scattering by longitudinally polarized two-magnons excitations. 

The former case is very well known in one dimensional systems \cite{sachdev}. Here the so-called spinons, which can be seen as domain walls, propagate freely and separate two "N\'eel" configurations \cite{kcuf3,lake}. In two and three dimensions, the situation is more complex as the interactions create an attractive potential between spinons that bind them to form conventional magnons \cite{piazza}. The quasi-degeneracy of the FM and AFM phases in the relevant part of the phase diagram phase for \ybti\, could cancel these interactions and release the spinons. In this view, the continuum would be characterized by the absence of edge singularity at its lower threshold, with the exception of the "on rod" Q positions, as revealed by the quasi-elastic like shape. Actually, classical examples of continuum demonstrate that both situations are possible: the classical De Cloizaux-Person spectrum in the case of the spin 1/2 Heisenberg chain, shows an edge singularity at the lower threshold \cite{cloizaux,sachdev}, the spectra decreasing roughly as $1/\omega$ (the exact exponent depends on the anisotropy \cite{schultz}). However, for ``XXZ'' chains, close to the Ising limit of a strong coupling along the z-axis, have a spectrum determined by gapped solitons, with no singularity at the edges \cite{ishimura}. Theoretical studies have shown that the essence of the singularity relies on the fact that many low-lying modes add up \cite{mikeska}. In this picture, recent theoretical work argue that the spontaneous magnetization along $\langle 0 0 1 \rangle$ below the critical temperature should confines the spinons by creating a string tension between them\cite{tchernyshyov}. This should result in the discretization of the two-spinon continuum into multiple spin-wave branches, a feature which is however not observed in the data, that especially show no significant evolution below $T_c$.

Alternatively, since a strongly reduced ordered momentum is the sign for significant non-linear longitudinal excitations, the scattering by a multi-magnons continuum should be considered. In this case, the ground state is long range ordered, with propagating magnons as the elementary (transverse) quasi-particles. In the longitudinal channel, however, fluctuations along the length of the ordered moment may develop \cite{christensen}. Experimental realizations are however quite scarce and usually give rise to a kind of tail above the spin wave dispersion which is still very well defined. For this reason, we believe that this second scenario is less relevant.

It should be stressed finally that the measurements performed at $T=4.5$~K (see Figure \ref{fig5b:TK}) show that the spectra tend to become more conventional when increasing the temperature. 
Further inelastic neutron scattering measurements should be performed as a function of temperature, to look for an eventual transition or cross-over between quantum and classical regimes, with spin dynamics respectively governed by quantum and thermal fluctuations. 
\section{Conclusion}

In this work, we show that the short range correlations that develop below 2~K in \ybti~ come along with non-conventional excitations: no well defined spin waves are observed in both the short range correlated and long range ordered regimes; rather, excitations spectra are characterized by a very broad and almost flat dynamical response which extends up to $1-1.5$~meV and does not evolve with temperature below 2~K, sometimes coexisting with a quasi-elastic response depending on the wave-vector.
Based on Monte carlo Spin Dynamics and RPA calculations, we determine a new set exchange couplings that allows to reproduce both the diffuse scattering in the short range correlated regime and the spin wave spectra observed in the field polarized phase. This careful determination of the exchange tensors places \ybti~ in the close vicinity to a FM/AFM phase boundary, and, in any case, far from the canonical quantum spin liquid state. This location points out the possible role of quantum fluctuations arising from these two competing phases.
We show further that conventional RPA fails to reproduce the experimental excitation spectra at low temperature. Spin dynamics simulations performed in the short range correlated regime reproduce however some features of the excitation spectra but lead to an energy bandwidth which is twice smaller than the experimental observations. 
We speculate that quantum fluctuations between FM and AFM phases govern the spin dynamics in \ybti\, and especially that the observed spectra rather correspond to a continuum of deconfined spinons as expected in quantum spin liquids, than simple paramagnons.

\acknowledgements

We acknowledge M. Gingras and L. Jaubert for fruitful discussions. We thank C. Paulsen for allowing us to use his SQUID dilution magnetometers. We also thank Ph. Boutrouille, (cryogeny group at LLB) for his technical help while setting up He dilution fridge.

\appendix

\section{Neutron data analysis}
\label{Annexe:neutron}

\begin{figure}[!t]
\centerline{
\includegraphics[keepaspectratio=1, width=8.5cm]{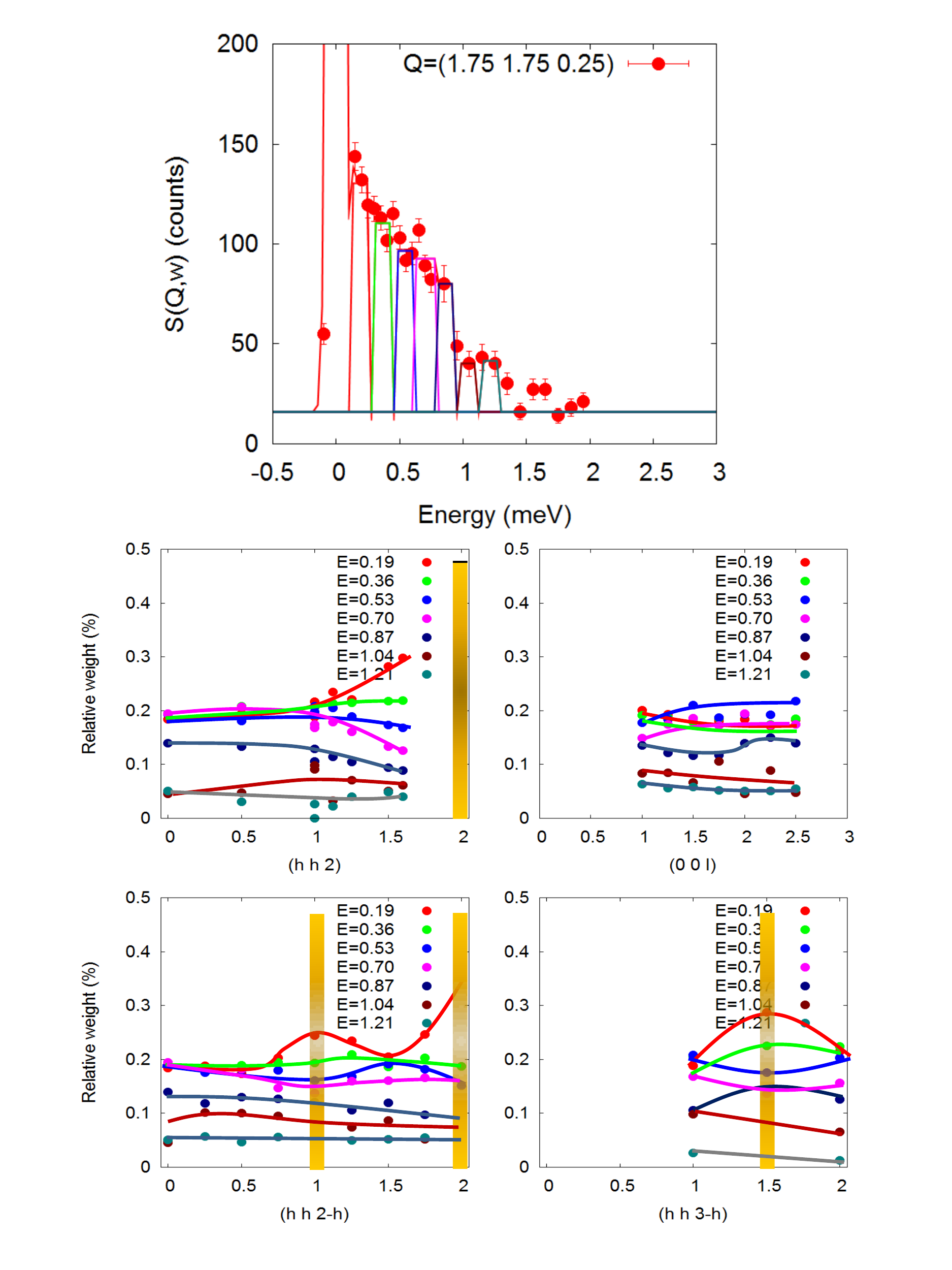}
}
\caption{(Color online) Top: example of raw data fitted according to the series of spectral bands $R_i(\omega)$. Bottom: Relative weights $A_i/\sum_j A_j$ of the different bands for the data recorded at base temperature along $(h~h~2)$, $(0~0~\ell)$, $(1~1~\ell)$ ,$(h~h~2-h)$ and $(h~h~3-h)$. $A_i$ is defined via the spectral function $(1+n(\omega)] \sum_{i=0,..,6} R_i(\omega)-R_i(-\omega)$, where $R_i(\omega)$ is constant, equal to $A_i$ in a given energy range centered at $E_{i=0,1,2,3,4,5,6} = 0.19, 0.36, 0.53, 0.70, 0.87, 1.04$ and $1.21$ meV, with a fixed width of $2 \Delta_r = 170$ $\mu$eV.}
\label{band}
\end{figure}

To further analyze the continuum of excitations observed in neutron experiments, the data can be modeled with the following function :
\begin{equation}
I(\omega) = C + G_r(\omega) + [1+n(\omega)] \sum_{i=0,..,6} R_i(\omega)-R_i(-\omega) 
\end{equation}
$C$ is a constant background, $G_r(\omega)$ a standard Gaussian profile of width $\Delta_r$ to model the incoherent elastic scattering at $\omega=0$, $1+n(\omega)$ is the detailed balance factor and $R_i(\omega)$ a series of flat spectral bands with a fixed width of $2 \Delta_r = 170$ $\mu$eV, and centered at $E_{i=0,1,2,3,4,5,6} = 0.19, 0.36, 0.53, 0.70, 0.87, 1,04$ and $1,21$ meV (see top of Figure \ref{band} for an example):
\begin{equation}
R_i(\omega) = A_i~\mbox{if}~|\omega-E_i| \leq \Delta_r~\mbox{and}~0~\mbox{elsewhere}
\end{equation}
The different spectra were analyzed through this model and the relative weights $A_i/\sum_j A_j$ are reported in Figure \ref{band} as a function of $\mathbf{Q}$ for the high symmetry directions. Those weights sequence in decreasing order as $\mathbf{Q}$ approaches the rods positions, close to $(2,2,2)$, $(1,1,1)$, $(3/2,3/2,3/2)$ and $(2,2,0)$ (see the orange bars), while the weights $i=0,1,2,3$ become roughly equal away from the rod. The spectra then appear "flat toped". Along $(00\ell)$ a weak maximum can be detected for the $i=2$ band ($E=0.53$ meV). 

These inelastic data, and especially the flat energy dependence demonstrates that on- and off-rods is a general feature of the spin dynamics throughout the Brillouin zone. Furthermore, the measurements show that the dynamical response extends up to $1-1.5$~meV.

\section{Connection between models}
\label{Annexe:conversion}

\subsection*{Model in terms of the full magnetic moment}

The widely accepted starting point is given by the following Hamiltonian
\begin{eqnarray}
H & = & H_{CEF} + H_{exc} + H_Z.
\label{gen_ham2}
\end{eqnarray}
$H_Z= g_J \sum_i \bm H \cdot \bm J_i$, is the Zeeman term, with $\bm H$ the applied magnetic field and $\bm J_i$ the magnetic moment at site $i$. $H_{CEF}$ is the crystal electric field (CEF) Hamiltonian and $H_{exc} = \sum_{ij} \bm J_i \tilde{\mathcal{K}_{ij}} \bm J_j$ is a bilinear coupling Hamiltonian, where the interaction tensor $\tilde{\mathcal{K}_{ij}}$ couples next neighbors magnetic moments $\bm J$ at sites $i$ and $j$. By symmetry arguments, the 9 coupling constants of the $3 \times 3$ tensor $\tilde{\mathcal{K}_{ij}}$ are reduced to only 4 \cite{rossprx}. Here, we assume an exchange tensor which is diagonal in the ($\mathbf{a},\mathbf{b},\mathbf{c})$ frame linked with a R-R bond \cite{malkin04}:
\begin{eqnarray*}
\mathbf{J}_i {\cal \tilde K} \mathbf{J}_j &=& \sum_{\mu,\nu=x,y,z} J_i^{\mu} 
\left( 
{\cal K}_a a_{ij}^{\mu} a_{ij}^{\nu} + 
{\cal K}_b b_{ij}^{\mu} b_{ij}^{\nu} + {\cal K}_c c_{ij}^{\mu} c_{ij}^{\nu} 
\right) J_j^{\nu} \\
& &- {\cal K}_4 \sqrt{2}~\vec{b}_{ij}.(\vec{J}_i \times \vec{J}_j)
\end{eqnarray*}
Alternatively, we can use ${\cal K}_{1,2,3,4}$ defined by the simple relations:
\begin{eqnarray*}
{\cal K}_1& = & \frac{{\cal K}_a + {\cal K}_c}{2}\\
{\cal K}_2& = & {\cal K}_b\\
{\cal K}_3& = & \frac{{\cal K}_a - {\cal K}_c}{2}
\end{eqnarray*}

\subsection*{Model in terms of a pseudo spin 1/2}

\begin{figure}[!t]
\centerline{
\includegraphics[trim=80px 0 70px 0, width=8.5cm, clip]{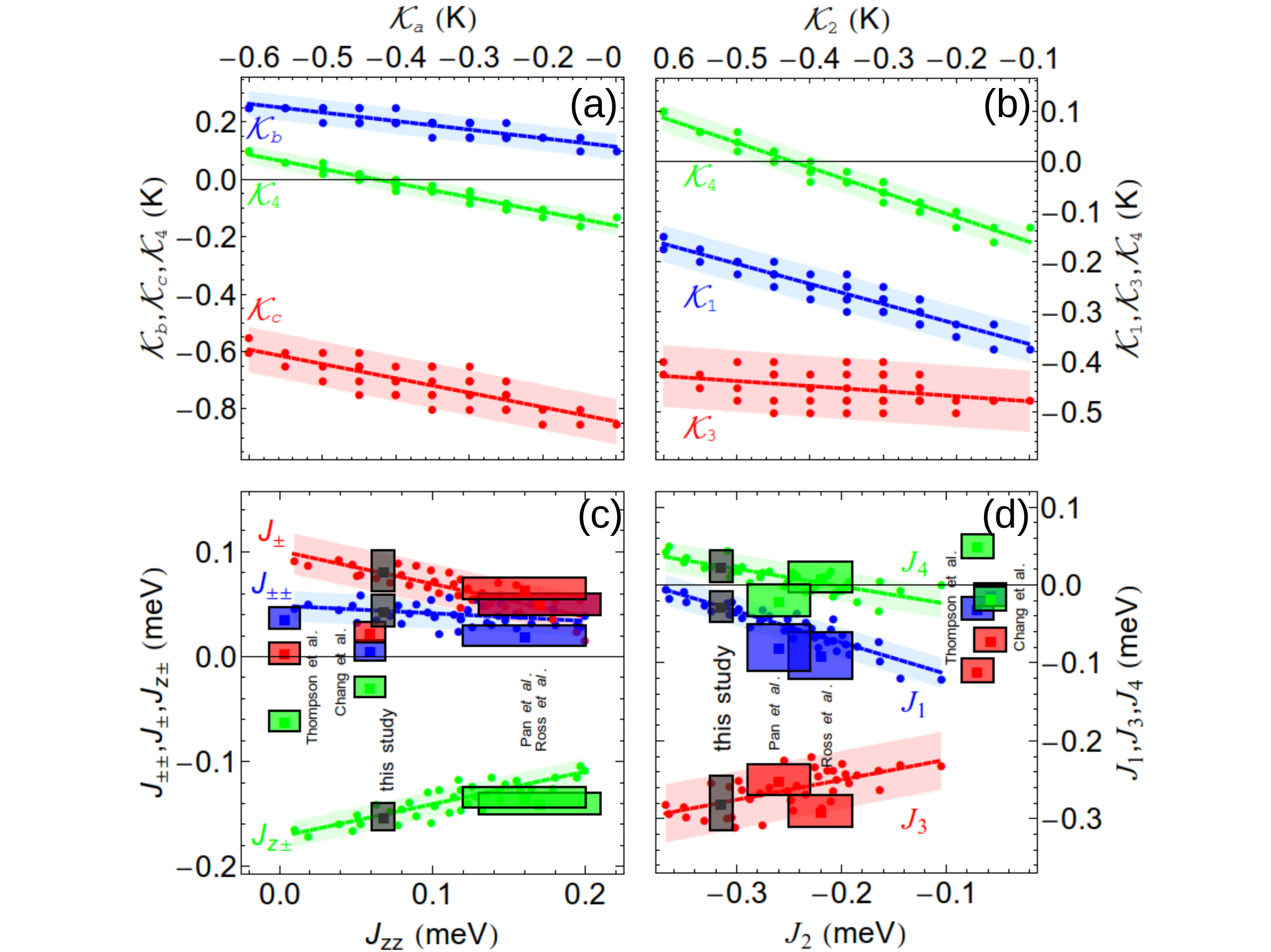}
}
\caption{(Color online) Correspondence between the different exchange coupling sets. The green, red and blue shaded areas correspond to the range of parameters giving a good agreement with the field induced spin wave spectra (see section \ref{ssec:parameters}).}
\label{corres}
\end{figure}

Since the energy gap between the CEF ground-state and the first excited levels are order of magnitudes larger than the exchange interactions and the Zeeman term, it is possible to define effective spin $1/2$ operators, denoted by $\mathbf{S}_i$, by projecting the full moment $\mathbf{J}_i$ onto the CEF ground-state doublet. The effective spin 1/2 Hamiltonian reads:
\begin{eqnarray}
H_{\mbox{eff}} & = & \sum_{ij} \mathbf{S}_i \tilde{J_{ij}} \mathbf{S}_j,
\label{eq:effective2}
\end{eqnarray}
A popular convention consists in using $({\sf J}_{\pm\pm},{\sf J}_{\pm},{\sf J}_{z\pm},{\sf J}_{zz})$ defined as:
\begin{eqnarray*}
H_{\mbox{eff}} &=& \sum_{i,j} {\sf J}_{zz} {\sf S}^z_i {\sf S}^z_j - {\sf J}_{\pm} \left({\sf S}^+_i {\sf S}^-_j + {\sf S}^-_i {\sf S}^+_j \right) \\
& &
+ {\sf J}_{\pm\pm} \left(\gamma_{ij} {\sf S}^+_i {\sf S}^+_j + \gamma^*_{ij} {\sf S}^-_i {\sf S}^-_j \right) \\
& &
+ {\sf J}_{z \pm} \left[ {\sf S}_i^z \left( \zeta_{ij} {\sf S}^+_j + \zeta^*_{ij} {\sf S}^-_j\right) + i \leftrightarrow j \right] 
\end{eqnarray*} 
where $\gamma_{ij}, \zeta_{ij}$ are numbers defined in Ref \onlinecite{rossprx,savary,wong}. Note that "sanserif" notations refer to local bases. The connection between ${\cal \tilde K}$ and ${\tilde J}$ is realized using the anisotropic Land\'e factor $g_{\perp,z}$ \cite{ersn}. Defining $\lambda_{\perp,z}=\frac{g_{\perp,z}}{g_J}$:
\begin{eqnarray*}
{\sf J}_{zz}  & = & \lambda_z^2 ~\frac{{\cal K}_a-2{\cal K}_c-4{\cal K}_4}{3} \\
{\sf J}_{\pm} & = & -\lambda_{\perp}^2 ~\frac{2{\cal K}_a-3{\cal K}_b-{\cal K}_c+4{\cal K}_4}{12} \\
{\sf J}_{z\pm} & = & \lambda_{\perp}~\lambda_z ~\frac{{\cal K}_a+{\cal K}_c-{\cal K}_4}{3 \sqrt{2}} \\
{\sf J}_{\pm \pm} & = & \lambda_{\perp}^2 ~\frac{2{\cal K}_a+3{\cal K}_b-{\cal K}_c+4{\cal K}_4}{12}
\end{eqnarray*}
Alternatively, a set of effective parameters $J_{1,2,3,4}$ can be used \cite{rossprx,ludo}: 
\begin{eqnarray*}
J_{1} & = & \frac{1}{3} \left( 2{\sf J}_{\pm \pm} + 4 {\sf J}_{\pm} + 2\sqrt{2} {\sf J}_{z\pm} - {\sf J}_{zz} \right) \\
J_{2} & = & \frac{1}{3} \left( 4{\sf J}_{\pm \pm} - 4 {\sf J}_{\pm} + 4\sqrt{2} {\sf J}_{z\pm} + {\sf J}_{zz} \right) \\
J_{3} & = & \frac{1}{3} \left(-4{\sf J}_{\pm \pm} - 2 {\sf J}_{\pm} + 2\sqrt{2} {\sf J}_{z\pm} - {\sf J}_{zz} \right) \\
J_{4} & = & \frac{1}{3} \left(2{\sf J}_{\pm \pm} - 2 {\sf J}_{\pm} - 2\sqrt{2} {\sf J}_{z\pm} - {\sf J}_{zz} \right)
\end{eqnarray*}
Figure \ref{corres} provides the different sets of parameters and correspondence between these 4 conventions able to capture the spin wave spectra measured in the field polarized phase of \ybti.
The correspondence between the optimal sets, obtained by taking into account the fit of the zero field diffuse scattering, is finally given in Table \ref{param}. 
\begin{table}
\begin{tabular}{cccc}
\hline
Full magnetic moment & Pseudo spin & Pseudo spin \\
(K) & (meV) & (meV) \\
\hline
${\cal K}_a=-0.48$ & ${\sf J}_{\pm\pm}=0.04$ & $J_{1}=-0.03$ \\
${\cal K}_b=0.23$ & ${\sf J}_{\pm}=0.085$   & $J_{2}=-0.32$ \\
${\cal K}_c=-0.67$ & ${\sf J}_{z\pm}=-0.15$   & $J_{3}=-0.28$ \\
${\cal K}_4=0.02$   & ${\sf J}_{zz}=0.07$      & $J_{4}=0.02$ \\
\hline
\end{tabular}
\caption{Optimal sets of exchange in \ybti}
\label{param}
\end{table}

\section{Classical spin dynamics simulations}
\label{Annexe:SpinDynamics}

We consider the Heisenberg model $\mathcal{H} = −\sum_{\langle ij \rangle} \mathbf{S}_i \tilde{J_{ij}} \mathbf{S}_j$, where the summation is limited to nearest neighbors, $\tilde{J_{ij}}$ is the interaction tensor, and $|\mathbf{S}_i|=1/2$ are classical pseudo-spins located at the pyrochlore sites. Our interest lies in the time evolution of the spin-pair correlations emerging in such a model. It is convenient to probe such dynamical correlations by calculating the dynamical scattering function $S(\bm Q, \omega)$, which can be done by combining Monte Carlo and spin dynamics methods.



The classical dynamics of the pseudo-spins is described by the non-linear Bloch equations 
\begin{eqnarray}
\frac{d \mathbf{S}_i}{dt} & = & \mathbf{S}_i \times \sum_{j} \tilde{J_{ij}} \mathbf{S}_j
\end{eqnarray}
where sites $j$ are the nearest neighbors of $i$. These equations of motions were numerically integrated using an 8th-order explicit Runge-Kutta (RK) method with an adaptive step-size control, offering an excellent compromise between accuracy and computation time.

The initial spin configurations used for the numerical integration are generated at each temperature by a hybrid Monte Carlo method using a single-spin-flip Metropolis algorithm \cite{juju,juju2}.
A thousand spin configurations are used at each temperature to evaluate the ensemble average in $S(\bm Q, \omega)$ while the number of Monte Carlo steps needed for decorrelation is adapted in such a way that the stochastic correlation between spin configurations is lower than $0.1$. The numerical results have been obtained for different lattice sizes ranging from $L=8$ ($N_S=8192$) to $L=16$ ($N_S=65536$) with periodic boundary conditions.

The classical Monte Carlo approach described above was also used to derive thermodynamics quantities such as the specific heat and the magnetic susceptibility, which allowed us to determine the phase diagram presented in Figure \ref{fig7:phasediag}.




\begin{thebibliography}{99}
\bibitem{footnote}Although a finite size scaling could allow to determine more precisely the critical temperatures, our goal here is only to qualitative discuss the possible consequences of lying close to the FM/AFM phase boundary.
\end{thebibliography}
\end{document}